\documentclass{article}

\usepackage{arxiv}

\usepackage[utf8]{inputenc} 
\usepackage[T1]{fontenc}    
\usepackage{url}            
\usepackage{booktabs}       
\usepackage{nicefrac}       
\usepackage{microtype}      
\usepackage{lipsum}
\usepackage[square,comma,numbers]{natbib}
\usepackage{amsfonts}
\usepackage{amsmath}
\usepackage{graphicx}
\usepackage{dcolumn}
\usepackage{soul}
\usepackage{bm}
\usepackage{overpic}
\usepackage{booktabs}
\usepackage{braket}
\usepackage{color}
\usepackage{multirow}
\usepackage{hyperref}
\title{Temporal similarity metrics for latent network reconstruction: The role of time-lag decay}

\author{
  Hao Liao \\
  National Engineering Laboratory for Big Data System Computing Technology \\Guangdong Province Key Laboratory of Popular High Performance Computers\\ College of Computer Science and Software Engineering \\
  Shenzhen University\\
  Shenzhen 518060, PR China \\
   \And
 Ming-Kai Liu \\
  National Engineering Laboratory for Big Data System Computing Technology \\Guangdong Province Key Laboratory of Popular High Performance Computers\\ College of Computer Science and Software Engineering \\
  Shenzhen University\\
  Shenzhen 518060, PR China \\
  \And
  Manuel Sebastian Mariani \\
  Institute of Fundamental and Frontier Sciences\\
  University of Electronic Science and Technology of China\\
  Chengdu 610051, PR China\\
  URPP Social Networks\\Universit$\ddot{a}$t Z$\ddot{u}$rich\\CH-8050 Switzerland\\
  \And
  Mingyang Zhou \\
  National Engineering Laboratory for Big Data System Computing Technology \\Guangdong Province Key Laboratory of Popular High Performance Computers\\ College of Computer Science and Software Engineering \\
  Shenzhen University\\
  Shenzhen 518060, PR China \\
  \And
  Xingtong Wu \\
  National Engineering Laboratory for Big Data System Computing Technology \\Guangdong Province Key Laboratory of Popular High Performance Computers\\ College of Computer Science and Software Engineering \\
  Shenzhen University\\
  Shenzhen 518060, PR China \\
}

\begin{document}
\maketitle

\begin{abstract}
When investigating the spreading of a piece of information or the diffusion of an innovation, we often lack information on the underlying propagation network.
Reconstructing the hidden propagation paths based on the observed diffusion process is a challenging problem which has recently attracted attention from diverse research fields.
To address this reconstruction problem, based on static similarity metrics commonly used in the link prediction literature, we introduce new node-node temporal similarity metrics. The new metrics take as input the time-series of multiple independent spreading processes, based on the hypothesis that two nodes are more likely to be connected if they were often infected at similar points in time. This hypothesis is implemented by introducing a time-lag function which penalizes distant infection times. We find that the choice of this time-lag strongly affects the metrics' reconstruction accuracy, depending on the network's clustering coefficient and we provide an extensive comparative analysis of static and temporal similarity metrics for network reconstruction. Our findings shed new light on the notion of similarity between pairs of nodes in complex networks.
\end{abstract}


\section{Introduction}
Our understanding of social networks is affected by the fact that, typically, we only have incomplete knowledge about the topology of real networks~\citep{clauset2008hierarchical,bakshy2012role}.
Aimed at overcoming this shortcoming, the problem of reconstructing missing links has attracted enormous attention from scholars from diverse fields (see ~\citep{martinez2017survey} for a recent review on the problem). Existing approaches to the network reconstruction problem include the use of local structural metrics~\citep{leicht2006vertex,martinez2017survey,daminelli2015common}, global walk-counting methods~\citep{katz1953new,martinez2017survey}, stochastic block models~\citep{guimera2009missing}, fitness-based methods~\citep{cimini2015systemic}, structural perturbation analysis~\citep{lu2015toward,liao2017ranking}, machine-learning techniques~\citep{grover2016node2vec}, among many others. Scholars have aimed to identify missing connections in a wide variety of systems, including protein-protein interaction networks~\citep{kovacs2018network}, neural networks~\citep{cannistraci2013link}, citation networks~\citep{ciotti2016homophily}, and social networks~\citep{liben2007link,backstrom2011supervised}.

In parallel, there has been recent interest~\citep{myers2010convexity,papadopoulos2012popularity,gomez2012inferring,zeng2013inferring,
shen2014reconstructing,liao2015reconstructing} on a different problem of network reconstruction: if we are only provided with information on the outcome of a dynamical process on an unknown propagation network, can we reconstruct the propagation network?
The problem -- which has been referred to as \emph{latent network reconstruction}~\citep{myers2010convexity} -- can be included in the broader class of problems that aim to reconstruct the properties of a spreading process (for instance, the seed node~\citep{brockmann2013hidden} or the epidemic parameters~\citep{myers2010convexity}) from data on observed realizations of the process. The question is fundamentally different from the traditional link prediction problem~\citep{martinez2017survey}: while link prediction studies~\citep{martinez2017survey} typically assume that only part of the network is hidden and needs to be reconstructed, here we assume that the topology of the propagation network is completely hidden.
The reconstruction problem studied here is important as we often deal with datasets where the propagation network is largely unknown: for instance, the owners of an online e-commerce platform might have complete information on the time-series of users' purchases, but lack information about the social connections between the users which might have affected, to some extent, the observed purchasing patterns.

Existing works have tackled the latent network reconstruction problem from various perspectives.
Among the most relevant contributions, Myers et al.~\cite{myers2010convexity} addressed the problem through a maximum-likelihood estimation method based on a cascade spreading model, which was further mapped into a convex optimization problem.
Gomez-Rodriguez et al.~\citep{gomez2012inferring} developed a faster maximum-likelihood method based on a cascade propagation model.
Shen et al.~\citep{shen2014reconstructing} leveraged compressed sensing theory to map the network reconstruction problem into a convex optimization problem.
Such a mapping is non-trivial and model-specific; they solved the problem for the Susceptible-Infected-Susceptible (SIS) and the "contact process" dynamics~\citep{shen2014reconstructing}.
The main limitation of these approaches is that they are model-dependent:
Different spreading models require the solution of a different set of equations. For example, in the compressed-sensing theory approach, the convex-optimization equations for the SIS and the contact process model differ substantially~\cite{shen2014reconstructing}.
Besides, the compressed-sensing approach to network reconstruction can be only applied to sparse networks~\cite{shen2014reconstructing}.

On the other hand, other studies~\cite{zeng2013inferring,liao2015reconstructing} have tackled the latent network reconstruction problem by means of simple similarity metrics.
With respect to convex optimization~\citep{myers2010convexity} and methods based on compressed sensing theory~\citep{shen2014reconstructing}, similarity metrics
have two main advantages: (1) They do not depend on the specific spreading model considered; (2) Their implementation is faster.
Temporal similarity metrics for the latent network reconstruction~\citep{liao2015reconstructing} build on the hypothesis that two nodes are more likely to be connected if independent spreading processes tend to infect them at similar times. A simple way to implement this assumption is to impose, for each pair $(i,j)$ of nodes that are infected by the same spreading process, a contribution to their similarity $s_{ij}$ in the form of a power-law decreasing function of the time lag between the two infection times~\citep{liao2015reconstructing}. For this reason, we refer to these metrics as temporal similarities with \emph{power-law time lag decay}.

Here, we develop new temporal similarity indexes based on the hypothesis that two nodes are more likely to be connected if independent spreading processes tend to infect them at two consecutive time steps of the dynamics. We refer to the new metrics as temporal similarities with \emph{one-step time lag decay}. Based on the power-law and one-step decay functions, for each of the eight classes of structural similarity metrics considered here, we construct two corresponding temporal similarity metrics. We compare their performance in reconstructing the whole propagation network in both synthetic and real data.
By analyzing $40$ empirical networks, we provide the first systematic performance comparison of temporal similarity metrics based on different classes of structural similarity metrics.

We find that for the Susceptible-Infected-Recovered (SIR) spreading dynamics~\citep{pastor2015epidemic}, for almost all the analyzed networks, the temporal similarities with one-step time lag decay outperform the temporal similarities with a power-law time lag decay. The performance gap is substantially larger for spreading processes sufficiently above their critical point. Besides, among all the classes of similarity metrics considered, we find that the temporal similarity metric with one-step time-lag decay based on the Cosine similarity~\citep{Salton1986} tends to outperform the other metrics; other competitive classes of similarity are the temporal variants of the Sorensen index~\citep{sorensen1948method} and the Jaccard similarity~\citep{jaccard1901etude}.
Results for two additional spreading models (Susceptible-Infected, SI, and Linear Threshold Model, LTM) are in qualitative agreement.

Our findings move the first steps toward an extensive benchmarking of methods for the reconstruction of a hidden topology from the available event time-series of a spreading process. Our work sheds new light on the notion of node-similarity based on the outcome of dynamical processes on networks, and it has potential implications for social network analysis that will be outlined in the Discussion section.

\begin{figure*}[t]
  \centering
  \includegraphics[width=17cm,scale=0.5]{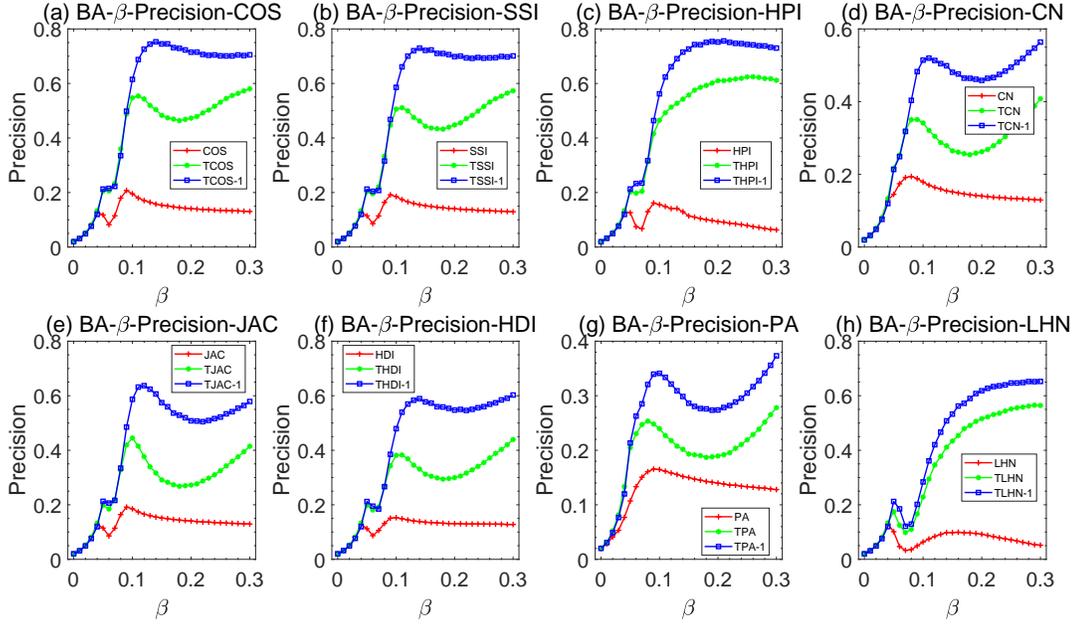}
\caption{Reconstruction precision of different similarity metrics as a function of $\beta$ for eight classes of similarity metrics (CN, COS, SSI, HPI, JAC, HDI, PA, LHN), for the SIR dynamics ($f=0.5$) on BA networks ($N=500$, $\langle k\rangle=5$).
The results are averaged over $50$ independent realizations. For sufficiently large $\beta$ values, temporal similarity metrics with one-step time-lag decay substantially outperform temporal similarity metrics with power-law time-lag and static metrics.
}
\label{fig1}
\end{figure*}

\begin{figure*}[htb]
  \centering
  \includegraphics[width=17cm,scale=0.1]{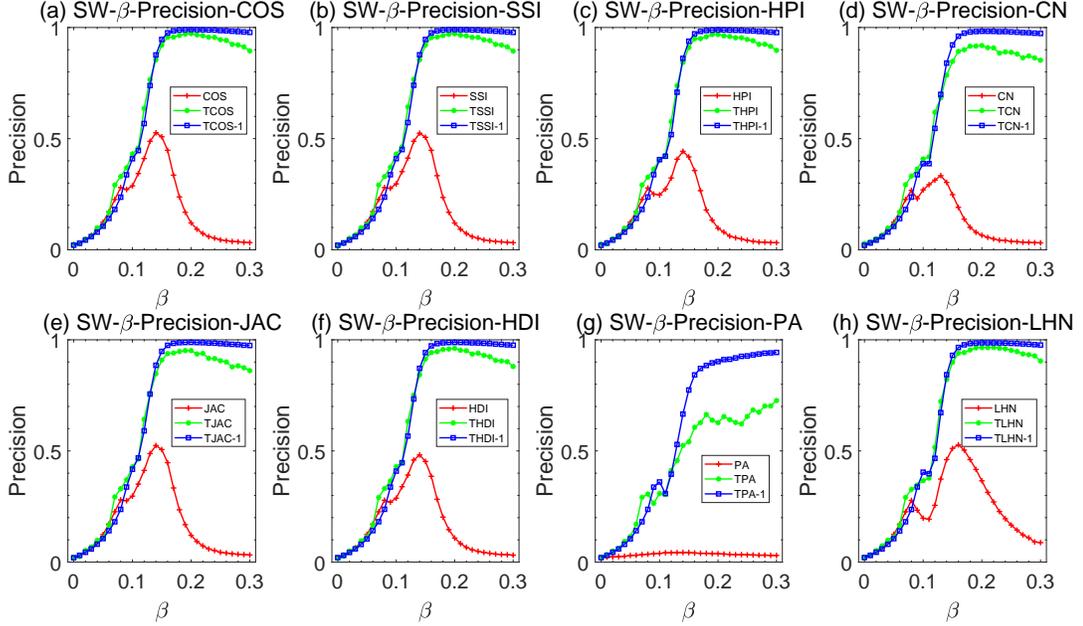}
\caption{(Color online) Reconstruction precision of different similarity metrics as a function of $\beta$ for eight classes of similarity metrics (CN, COS, SSI, HPI, JAC, HDI, PA, LHN), for the SIR dynamics ($f=0.5$) on SW networks ($N=500$, $P = 0.1$, $\langle k\rangle=5$). The results are averaged over $50$ independent realizations. For sufficiently large $\beta$ values, temporal similarity metrics with one-step time-lag decay substantially outperform temporal similarity metrics with power-law time-lag and static metrics.
}
\label{fig2}
\end{figure*}

\section{Results}
\subsection{Problem statement}

We assume that there is a unipartite network (whose adjacency matrix is denoted by $\mathsf{A}$) whose topology is unknown, and our goal is to reconstruct it. Our available information is the time-stamped list of adoptions of multiple items that diffuse through a given spreading process. Entry $(i,\alpha,t_{i\alpha})$ in this list tells us that node $i$ adopted item $\alpha$ at time $t_{i\alpha}$. The adoption processes considered here is ruled by the SIR dynamics~\citep{pastor2015epidemic}: the "adoption" of item $\alpha$ corresponds to the "infection" during realization $\alpha$ of the SIR dynamics. For this reason, in the following, we will use "adoption" and "infection" interchangeably.

We consider $40$ empirical unipartite networks; among these, $20$ are information networks (details in the Supplementary Material). We generate the time-series $\{(i,\alpha,t_{i\alpha})\}$ of adoptions by running, for each network, $50$ independent realizations of the SIR spreading dynamics initiated by a fraction $f$ of initiators (see Methods for details)~\citep{liao2015reconstructing}. Each independent realization $\alpha$ of the spreading process is therefore interpreted as an item that gradually diffuses across the network. In fact, the time-series $\{(i,\alpha,t_{i\alpha})\}$ can be interpreted as a temporal bipartite network~\citep{holme2012temporal}; we denote by $\mathsf{R}$ the incidence matrix of the corresponding time-aggregate bipartite network: $R_{i\alpha}=1$ if node $i$ adopted item $\alpha$.

We address the following problem. Assuming that we only know $\{i,\alpha, t_{i\alpha}\}$, which is the best method to reconstruct the $E$ edges of $\mathsf{A}$ from $\{i,\alpha, t_{i\alpha}\}$? While, in principle, several techniques of network reconstruction can be designed~\citep{martinez2017survey,shen2014reconstructing,liao2015reconstructing}, we narrow our focus to similarity metrics that aim to infer the similarity $s_{ij}$ of two nodes $i$ and $j$ based on their co-adoption patterns~\citep{zeng2013inferring,liao2015reconstructing}.
The definitions of the metrics of interest are provided in Sections~\ref{sec:metrics1} and~\ref{sec:metrics2}.

Such similarity metrics produce a ranking of the pairs of nodes (potential edges) in descending order of $s_{ij}$.
Assuming that we know the number of edges $E$ of the underlying propagation network $\mathsf{A}$, the $E$ top-ranked links by $s_{ij}$ form the network $\mathsf{A}^{(s)}$ reconstructed by metric $s$. It is natural to assess the precision of the metric $s_{ij}$ by measuring the fraction of common links between $\mathsf{A}$ and $\mathsf{A}^{(s)}$. This metric is typically referred to as precision in the link prediction~\citep{martinez2017survey} and information filtering literature, and we use it to evaluate the reconstruction performance of the similarity metrics.
The results for another evaluation metric\footnote{Differently from the precision metric, the AUC metric is independent of $E$.} (Area Under the Curve, AUC~\cite{lu2011link}) are in qualitative agreement with those obtained with the precision (Figs. S8).

\begin{figure*}[htb]
  \centering
  \includegraphics[width=17cm,scale=0.5]{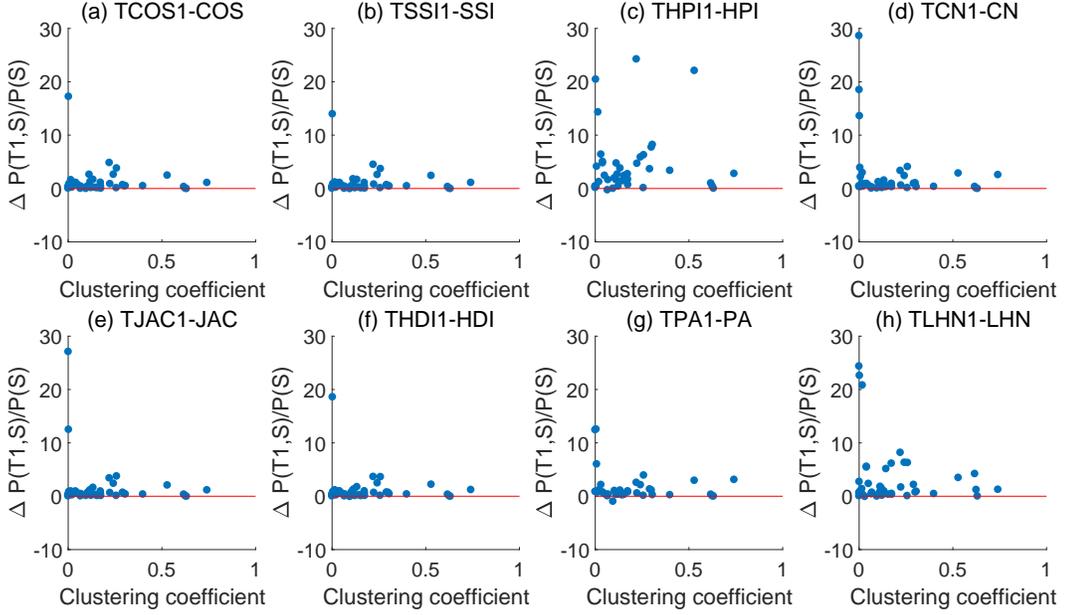}
\caption{Reconstruction precision relative difference $\Delta P(T1,S)/P(S)=(P(T1)-P(S))/P(S)$ as a function of the network clustering coefficient. Each dot represents an empirical network; we analyzed $40$ empirical contact networks. For all classes of similarity, almost all the empirical networks fall above the $P(T1)=P(S)$ red line. We use $\beta = 4 \beta_c$ and $f = 0.5$ here. The results are averaged over $50$ independent realizations.
}
\label{fig3}
\end{figure*}
\begin{figure*}[t]
  \centering
  \includegraphics[width=17cm,scale=0.5]{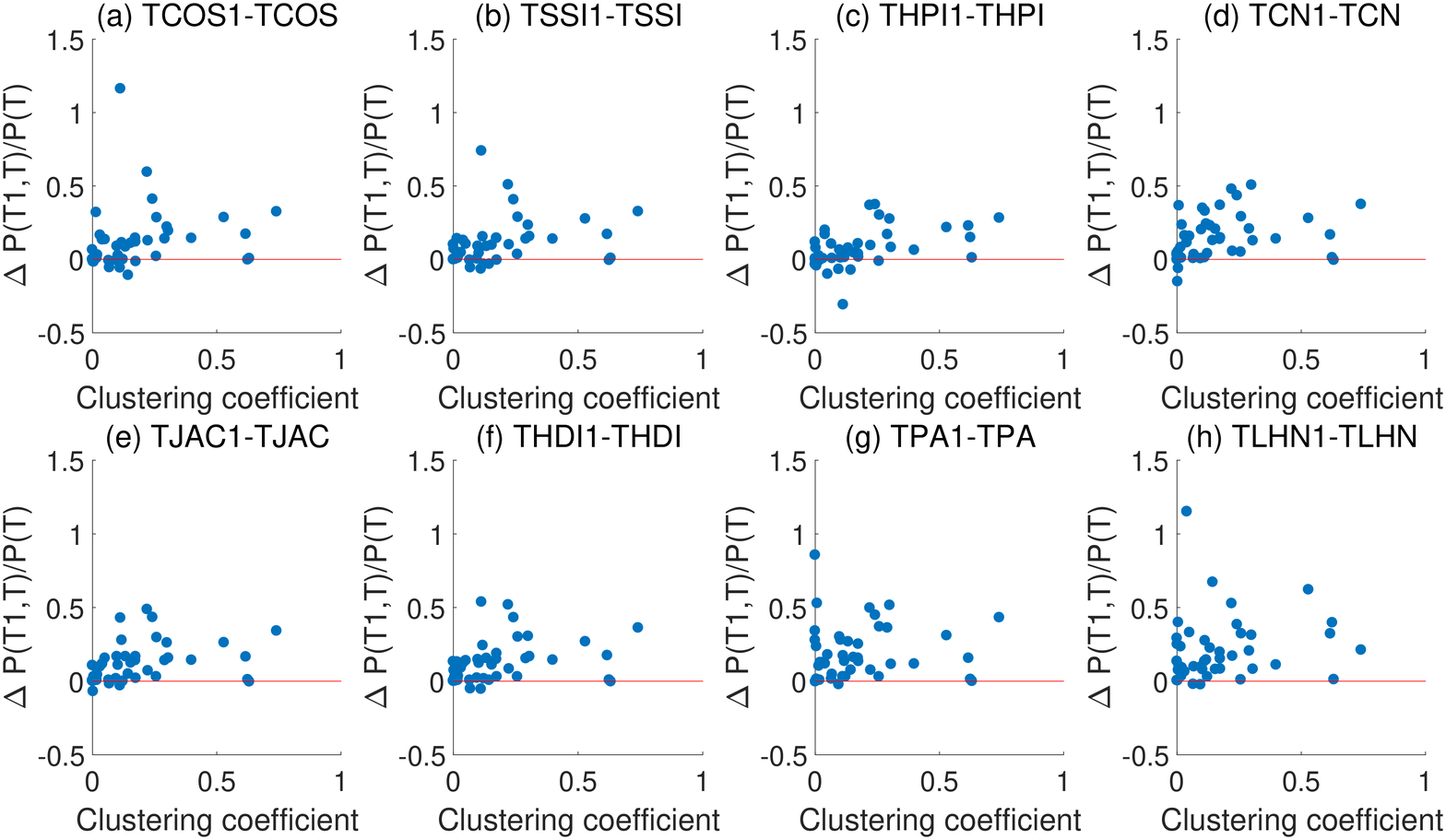}
\caption{Reconstruction precision relative difference $\Delta P(T1,T)/P(T)=(P(T1)-P(T))/P(T)$ as a function of the network clustering coefficient. Each dot represents an empirical network; we analyzed $40$ empirical contact networks. For all classes of similarity, almost all the empirical networks fall above the $P(T1)=P(T)$ red line; the only exceptions are some of the networks with low clustering coefficient. We use $\beta = 4 \beta_c$ and $f = 0.5$ here. The results are averaged over $50$ independent realizations.
}
\label{fig4}
\end{figure*}

\subsection{From structural to temporal similarity metrics}
\label{sec:metrics1}

We consider here eight classes of structural similarities~\citep{martinez2017survey}:
common neighbors (CN), Jaccard Index (Jac), Leicht-Holme-Newman Index (LHN), Cosine Index (COS), Sorensen Index (SSI), Hub Promoted Index (HPI), Hub Depressed Index (HDI), Preferential Attachment (PA).
These structural metrics have been used by researchers from diverse domains to address various problems in network analysis. They have been applied to the reconstruction of missing links in networks where only a part of the topology is available~\cite{clauset2008hierarchical,lu2011link}, to the prediction of new connections in social and information systems~\cite{liben2007link}, and to the latent network reconstruction problem studied here as well~\cite{zeng2013inferring}.

For each class\footnote{X is a placeholder here. E.g., X can represent common neighbors CN.} X of similarities, we consider the standard static metric~\citep{martinez2017survey} (directly denoted as X), and two \emph{temporal} similarity metrics: temporal metrics with the power-law time-lag decay (denoted as TX)~\citep{liao2015reconstructing}, and the new temporal metrics with the one-step time-lag decay (denoted as TX1).
The last two classes of metrics differ in how the similarity score of a given pair $(i,j)$ of nodes depends on the \emph{time lag} $t_{i\alpha}-t_{j\alpha}$ between node $i$'s and $j$'s adoption times $t_{i\alpha}$ and $t_{j\alpha}$ for item $\alpha$. We refer to the Methods section for all the definitions.

To illustrate the main idea behind each class of metrics, we define here the common-neighbors metrics: static common neighbors (CN), temporal common neighbors with a power-law decay of time-lag (TCN), and temporal common neighbors with one-step decay of time lag (TCN1).
The common neighbors (CN) of a given pair $(i,j)$ of nodes is simply given by~\citep{martinez2017survey}
\begin{equation}
s_{ij}^{CN}=\sum_{\alpha}R_{i\alpha}R_{j\alpha}.
\end{equation}
According to this definition, two nodes are similar (and, therefore, more likely to be connected in the hidden unipartite network) if they often adopted the same item.

Zeng~\citep{zeng2013inferring} found that this metric and similar \emph{static} metrics can be used to reconstruct the topology of a hidden network based on the time-series of a spreading dynamics.
Subsequently, the static metric proved to be sub-optimal with respect to time-aware metrics~\citep{liao2015reconstructing}. Indeed, while it is plausible that two nodes that often adopt the same item at similar times are more likely to be connected, the same is not necessarily true if the common adoptions happen at very distant points in time: given two adopters $i$ and $j$, with $t_{i\alpha}\ll t_{j\alpha}$, item $\alpha$ might indeed have reached $j$ though a long network path, without the two nodes being directly connected.

To penalize longer time lags, ~\citep{liao2015reconstructing} introduced the \emph{temporal common neighbors with power-law time-lag decay} (TCN) as
\begin{equation}
s_{ij}^{TCN}=\sum_{\alpha}R_{i\alpha}\,R_{j\alpha}\,|t_{i\alpha}-t_{j\alpha}|^{-1}(1-\delta_{t_{i\alpha},t_{j\alpha}}).
\end{equation}

This time-aware metric significantly outperforms its static counterpart, $s^{CN}$, in the latent network reconstruction~\citep{liao2015reconstructing}.
However, as a consequence of the power-law function, the similarity $s^{TCN}$ of a given pair of nodes receives substantial non-zero contributions also when the two nodes adopt the same item at substantially different times.

\begin{figure*}[t]
  \centering
  \includegraphics[width=17cm,scale=0.5]{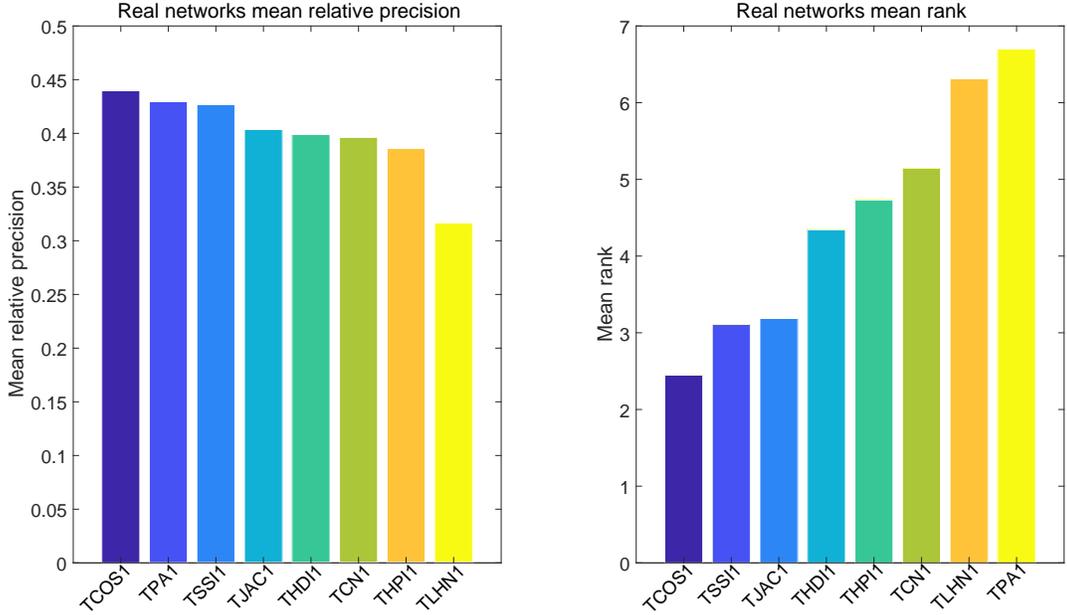}
\caption{Mean relative precision (higher values correspond to better performance) and mean rank (lower values correspond to better performance) of the eight TX1 metrics. According to both evaluation metrics, TCOS1 is the best-performing metric, followed by TSSI1 and TJAC1. We use $\beta = 4 \beta_c$ and $f = 0.5$ here. The results are averaged over $50$ independent realizations.
}
\label{fig5}
\end{figure*}

In this work, we introduce the \emph{temporal common neighbors with a one-step decay time-lag decay} (TCN1) as
\begin{equation}
s_{ij}^{TCN1}=\sum_{\alpha}R_{i\alpha}\,R_{j\alpha}\,\delta_{|t_{i\alpha}-t_{j\alpha}|,1}.
\end{equation}
According to this definition, the similarity $s^{TCN1}$ of a given pair $(i,j)$ of nodes only receives a contribution when the two nodes adopt the same item at two consecutive time steps.

Analogous definitions for the other seven classes of similarities and their temporal variants with power-law and one-step time-lag decay are provided in the Methods section. The goal of the rest of the paper is to extensively compare the performance of these metrics in reconstructing both synthetic and empirical networks.

\subsection{Reconstruction of synthetic networks}

We start our investigation from synthetic networks generated with the Barab\'asi-Albert model~\citep{barabasi1999emergence} (see Methods for the generation details). Fig.~\ref{fig1} shows our reconstruction results: each panel refers to a class of similarities; for each class of similarities (e.g., common neighbors), we show the results for the static metric (CN), the temporal metric with power-law time lag decay (TCN), and the new temporal metric with one-step time lag decay (TCN1). The precision values attained by the metrics are shown as a function of the transmission probability $\beta$ of the SIR spreading process.

For each considered structural metric (e.g., CN), for sufficiently large $\beta$ values, the corresponding temporal metric with one-step decay (TCN1) performs significantly better than the corresponding temporal metric with power-law decay (e.g., TCN).
As we reduce $\beta$, spreading processes tend to die out more rapidly, and it becomes increasingly harder to correctly reconstruct the underlying diffusion network; in the small-$\beta$ regime, the temporal metrics with a one-step and power-law decay perform similarly.
As expected~\citep{liao2015reconstructing}, the time-aware metrics significantly outperform the static metric.

Fig.~\ref{fig2} shows analogous results for a small-world network ~\citep{watts1998collective} (see Methods for the generation details). We observe again a systematic performance edge of the temporal metrics with a one-step time lag decay over the temporal metrics with power-law time lag decay, yet this gap is smaller than in the BA networks.

\subsection{Reconstruction of real networks}

Our results on synthetic networks suggest that the temporal metrics with one-step time lag decay reconstruct synthetic contact networks better than the temporal metrics with power-law time lag decay.
To further validate this assertion, we analyzed $40$ empirical contact networks of diverse nature including $20$ information networks (details in the Supplementary Material).

For almost all the analyzed datasets, the temporal metrics with one-step time lag decay substantially improve the reconstruction accuracy with respect to both static (Fig.~\ref{fig3}) and temporal metrics with power-law time lag decay (Fig.~\ref{fig4}). The only networks where the temporal metrics with power-law time-lag decay can outperform the temporal metrics with one-step time-lag decay are those with low clustering coefficient\footnote{In our work, we use the average local clustering coefficient as a metric for clustering. For each node $i$ in the network, we calculate the number $K_i$ of existing edges that connect nodes that are connected with $i$, and the maximum number $E_i$ of possible links between $i$'s neighbors. For an undirected graph, $E_i=k_i\,(k_i-1)/2$. Finally, we define $i$'s local clustering coefficient $C_i$ = $K_i/E_i$, and the network's clustering coefficient as $C=N^{-1}\sum_i C_i$. }. This is intuitive: In a network with lower clustering, it is less likely that two non-connected nodes are reached by long propagation paths. This mitigates the advantage of considering only adoptions with one-step time lag when computing the similarity score of a given pair of nodes.

The results in Fig.~\ref{fig4} were obtained with $\beta= 4 \,\beta_c$, where $\beta_c$ is the epidemic threshold~\citep{pastor2015epidemic}. As expected from the synthetic network analysis, we find that for larger $\beta$ values (Fig. S1), the one-step time lag metrics show better reconstruction accuracy for the vast majority of datasets and considered metrics. On the other hand, for lower $\beta$ values, there is not a clear advantage of the metrics with the one-step time-lag decay (Figs. S2-S3).

So far, we have compared similarities of the same class (e.g., common neighbors) with different time-lag decay functions. A natural question arises: what is the relative performance of the eight temporal metrics TX1 with one-step time-lag decay obtained from the eight different classes of similarities? We compare the eight metrics' performance across the $40$ empirical datasets considered here. We refer to Figs. S4-S5 for the results on individual datasets. To gain a general understanding of the metrics' performance, we aggregate the metrics' performance over the analyzed networks. To this end, we consider two evaluation metrics: the metrics' mean rank~\citep{muscoloni2017local} and the mean relative precision.

To compute the metrics' mean rank, for each dataset $d$, we rank the eight TX1 metrics in order of decreasing precision. We denote by $r_d(s)\in\{1,2,\dots,8\}$ the ranking position metric $s$ for dataset $d$.
Given $D$ analyzed empirical networks ($D=40$ in our work), the mean rank $\overline{r(s)}$ of metric $s$ is simply defined as $\overline{r(s)}=D^{-1}\sum_{d=1}^{D}r_d(s)$. Better performing metric should exhibit lower mean rank values~\citep{muscoloni2017local}.
In addition, denoting by $P_d(s)$ the precision achieved by metric $s$ in dataset $d$, we define the mean relative precision $\overline{P(s)}$ of metric $s$ as $D^{-1}\sum_{d=1}^{D}P_d(s)/\max_{s'}\{P_d(s')\}$. Better performing metric should exhibit larger mean relative precision values.

Both evaluation metrics lead to the same overall conclusion (Fig.~\ref{fig5}): on average, the TCOS1 (temporal Cosine with one-step time-lag decay) metric
\begin{equation}
s_{ij}^{TCOS1}=\frac{\sum_{\alpha } {R_{i\alpha}R_{j\alpha}\delta_{|t_{i\alpha}-t_{j\alpha}|,1}}}{\sqrt{\sum_{\alpha} R_{i\alpha}\sum_{\alpha} R_{j\alpha}}}
\end{equation}
is the best-performing metric, followed by TSS1 and TJAC1 (see Methods for their definition).
While TCOS1 provides us with a computationally fast metric to reconstruct the hidden topology, its mean precision is $0.349$.
This leaves the door open for future performance improvements, possibly based on new similarity metrics or more sophisticated methods.

\section{Discussion}

Our work provided a systematic benchmarking of temporal similarity metrics with respect to their accuracy in reconstructing a hidden network topology. The reconstruction was more accurate for SIR spreading processes with a large transmission probability, i.e., in the supercritical regime.
On both real and synthetic networks, we found that temporal metrics with one-step time-lag decay perform systematically better than metrics with power-law time-lag decay. Besides, we found that the temporal cosine metric with one-step time-lag decay is the best-performing metric.
Differently from maximum-likelihood methods~\citep{gomez2012inferring} and compressed-sensing theory approaches~\citep{myers2010convexity}, the temporal similarity metrics considered here are general and not restricted to a specific dynamics. In this sense, they can be interpreted not only as parsimonious and effective reconstruction tools, but also as general baselines against which more sophisticated, model-specific reconstruction techniques can be evaluated.
While we focused on the SIR dynamics, we also assessed the metrics' performance for two additional spreading models: the Susceptible-Infected (SI) model~\citep{anderson1992infectious,starnini2013immunization} and the Linear Threshold Model (LTM)~\cite{granovetter1978threshold,kempe2003maximizing,chen2010scalable}. The results obtained for these two models are in qualitative agreement with the results obtained for the SIR model~(Figs. S6-S7), supporting the generality of our conclusions.

Our study paves the way for several extensions.
Temporal similarity metrics might be applied to other network reconstruction problems, such as the problem where part of the topology is known~\cite{lu2011link} and the matching of user accounts across different domains or devices~\cite{chen2018effective,li2018matching}.
Even more intriguingly, one can attempt to reconstruct the hidden topology of a social network based on the observed dynamics of real diffusion processes. For instance, from the observed spreading dynamics of many pieces of information, one might attempt to reconstruct propagation networks in social media~\citep{pei2014searching} and e-commerce platforms~\citep{medo2016identification}. The results presented here support metrics based on one-step time lags as the best-performing ones in the latent network reconstruction task. While the time step of the dynamics is unambiguously defined for simulated processes, the same does not hold for real spreading processes. Using temporal similarity metrics to reconstruct propagation topologies based on real time-series data will likely require us to first identify the typical timescale needed for a given piece of information to be transmitted from an individual to another, and then to use this typical timescale as the time-lag parameter in the similarity metric.

Finally, our work contributes to the rich literature on similarity on social and information networks~\citep{li2011link,li2012robust,sun2016prediction,gong2016influence,chen2017reality,peng2017social}.
Previous research has stressed the role of structural similarity metrics, i.e., similarity metrics based either on the time-aggregate contact network of individuals (who is connected to whom)~~\citep{huang2009time,xia2015learning,lee2017measure} or on the time-aggregate user-item bipartite adoption network (who collected what)~~\citep{li2012robust,zeng2013inferring}.
Here, we combined structure and temporal information (who collected what at which time) to define temporal similarity metrics that are effective in the propagation network task. We envision that future research on social and information network analysis might further develop simple yet well-performing time-aware metrics for network reconstruction.

\section{Methods}

\subsection{Temporal similarity metrics}
\label{sec:metrics2}

For each class $C$ of similarity metrics, we define three metrics: a static metric $C$, a temporal metric with power-law time-lag decay $TC$, and a temporal metric with one-step time-lag decay $TC1$. In our work, we consider eight classes C of similarities: Common Neighbors (CN), Jaccard (Jac), Cosine (COS), Leicht-Holme-Newman (LHN), Sorensen Index (SSI), Hub-promoted Index (HPI), Preferential Attachment (PA), Hub-depressed Index (HDI).
As we already defined the three CN similarities in the main text, we define here the metrics based on the seven additional classes.

\paragraph*{Jaccard (Jac) similarity} We define three metrics:
\begin{itemize}
\item Jaccard similarity (Jac):
\begin{equation}
s_{ij}^{Jac}=\frac{\sum_{\alpha} R_{i\alpha}\,R_{j\alpha}}{\sum_{\alpha}(R_{i\alpha}+R_{j\alpha}-R_{i\alpha}\,R_{j\alpha})}.
\end{equation}
\item Temporal Jaccard similarity with power-law time-lag decay (TJac):
\begin{equation}
s_{ij}^{TJac}=\frac{\sum_{\alpha} R_{i\alpha}\,R_{j\alpha}\,|t_{i\alpha}-t_{j\alpha}|^{-1}(1-\delta_{t_{i\alpha},t_{j\alpha}})}{\sum_{\alpha}(R_{i\alpha}+R_{j\alpha}-R_{i\alpha}\,R_{j\alpha})}.
\end{equation}
\item Temporal Jaccard similarity with one-step time-lag decay (TJac1):
\begin{equation}
s_{ij}^{TJac1} =\frac{\sum_{\alpha}R_{i\alpha}R_{j\alpha}\delta_{|t_{i\alpha}-t_{j\alpha}|,1}}{\sum_{\alpha}(R_{i\alpha}+R_{j\alpha}-R_{i\alpha}R_{j\alpha})}.
\end{equation}
\end{itemize}

\paragraph*{Cosine (COS) similarity} We define three metrics:
\begin{itemize}
\item Cosine similarity (COS):
\begin{equation}
s_{ij}^{COS}=\frac{\sum_{\alpha } {R_{i\alpha}R_{j\alpha}}}{\sqrt{\sum_{\alpha} R_{i\alpha}\sum_{\alpha} R_{j\alpha}}}.
\end{equation}
\item Temporal Cosine similarity with power-law time-lag decay (TCOS):
\begin{equation}
s_{ij}^{TCOS}=\frac{\sum_{\alpha } {R_{i\alpha}R_{j\alpha}|t_{i\alpha}-t_{j\alpha}|^{-1}(1-\delta_{t_{i\alpha},t_{j\alpha}})}}{\sqrt{\sum_{\alpha} R_{i\alpha}\sum_{\alpha} R_{j\alpha}}}.
\end{equation}
\item Temporal Cosine similarity with one-step time-lag decay (TCOS1):
\begin{equation}
s_{ij}^{TCOS1}=\frac{\sum_{\alpha } {R_{i\alpha}R_{j\alpha}\delta_{|t_{i\alpha}-t_{j\alpha}|,1}}}{\sqrt{\sum_{\alpha} R_{i\alpha}\sum_{\alpha} R_{j\alpha}}}.
\end{equation}
\end{itemize}

\paragraph*{Leicht-Holme-Newman Index (LHN) similarity} We define three metrics:
\begin{itemize}
\item Leicht-Holme-Newman Index similarity (LHN):
\begin{equation}
s_{ij}^{LHN} =\frac{\sum_{\alpha}R_{i\alpha}R_{j\alpha}}{\sum_{\alpha}R_{i\alpha}\sum_{\alpha}R_{j\alpha}}.
\end{equation}
\item Temporal Leicht-Holme-Newman Index similarity with power-law time-lag decay (TLHN):
\begin{equation}
s_{ij}^{TLHN} =\frac{\sum_{\alpha}R_{i\alpha}R_{j\alpha}|t_{i\alpha}-t_{j\alpha}|^{-1}(1-\delta_{t_{i\alpha},t_{j\alpha}})}{\sum_{\alpha}R_{i\alpha}\sum_{\alpha}R_{j\alpha}}.
\end{equation}
\item Temporal Leicht-Holme-Newman Index similarity with one-step time-lag decay (TLHN1):
\begin{equation}
s_{ij}^{TLHN1} =\frac{\sum_{\alpha}R_{i\alpha}R_{j\alpha}\delta_{|t_{i\alpha}-t_{j\alpha}|,1}}{\sum_{\alpha}R_{i\alpha}\sum_{\alpha}R_{j\alpha}}.
\end{equation}
\end{itemize}

\paragraph*{S{\o}rensen Index (SSI) similarity} We define three metrics:
\begin{itemize}
\item S{\o}rensen Index similarity (SSI):
\begin{equation}
s_{ij}^{SSI}=\frac{ 2\times \sum_{\alpha }{R_{i\alpha}R_{j\alpha}}}{\sum_{\alpha} R_{i\alpha}+\sum_{\alpha} R_{j\alpha}}.
\end{equation}
\item Temporal S{\o}rensen Index similarity with power-law time-lag decay (TSSI):
\begin{equation}
s_{ij}^{TSSI}=\frac{ 2\times \sum_{\alpha } {R_{i\alpha}R_{j\alpha}|t_{i\alpha}-t_{j\alpha}|^{-1}(1-\delta_{t_{i\alpha},t_{j\alpha}})}}{\sum_{\alpha} R_{i\alpha}+\sum_{\alpha} R_{j\alpha}}.
\end{equation}
\item Temporal S{\o}rensen Index similarity with one-step time-lag decay (TSSI1):
\begin{equation}
s_{ij}^{TSSI1}=\frac{ 2\times \sum_{\alpha } {R_{i\alpha}R_{j\alpha}\delta_{|t_{i\alpha}-t_{j\alpha}|,1}}}{\sum_{\alpha} R_{i\alpha}+\sum_{\alpha} R_{j\alpha}}.
\end{equation}
\end{itemize}

\paragraph*{Hub Promoted Index (HPI) similarity} We define three metrics:
\begin{itemize}
\item Hub Promoted Index similarity (HPI):
\begin{equation}
s_{ij}^{HPI}=\frac{\sum_{\alpha}R_{i\alpha}R_{j\alpha}}{min\{\sum_{\alpha}R_{i\alpha},\sum_{\alpha}R_{j\alpha}\}}
\end{equation}
\item Temporal Hub Promoted Index similarity with power-law time-lag decay (THPI):
\begin{equation}
s_{ij}^{THPI}=\frac{\sum_{\alpha}R_{i\alpha}R_{j\alpha}|t_{i\alpha}-t_{j\alpha}|^{-1}(1-\delta_{t_{i\alpha},t_{j\alpha}})}{min\{\sum_{\alpha}R_{i\alpha},\sum_{\alpha}R_{j\alpha}\}}
\end{equation}
\item Temporal Hub Promoted Index similarity with one-step time-lag decay (THPI1):
\begin{equation}
s_{ij}^{THPI1}=\frac{\sum_{\alpha}R_{i\alpha}R_{j\alpha}\delta_{|t_{i\alpha}-t_{j\alpha}|,1}}{min\{\sum_{\alpha}R_{i\alpha},\sum_{\alpha}R_{j\alpha}\}}
\end{equation}
\end{itemize}

\paragraph*{Hub Depressed Index (HDI) similarity} We define three metrics:
\begin{itemize}
\item Hub Depressed Index similarity (HDI):
\begin{equation}
s_{ij}^{HDI}=\frac{\sum_{\alpha}R_{i\alpha}R_{j\alpha}}{max\{\sum_{\alpha}R_{i\alpha},\sum_{\alpha}R_{j\alpha}\}}
\end{equation}
\item Temporal Hub Depressed Index similarity with power-law time-lag decay (THDI):
\begin{equation}
s_{ij}^{THDI}=\frac{\sum_{\alpha}R_{i\alpha}R_{j\alpha}|t_{i\alpha}-t_{j\alpha}|^{-1}(1-\delta_{t_{i\alpha},t_{j\alpha}})}{max\{\sum_{\alpha}R_{i\alpha},\sum_{\alpha}R_{j\alpha}\}}
\end{equation}
\item Temporal Hub Depressed Index similarity with one-step time-lag decay (THDI1):
\begin{equation}
s_{ij}^{THDI1}=\frac{\sum_{\alpha}R_{i\alpha}R_{j\alpha}\delta_{|t_{i\alpha}-t_{j\alpha}|,1}}{max\{\sum_{\alpha}R_{i\alpha},\sum_{\alpha}R_{j\alpha}\}}
\end{equation}
\end{itemize}

\paragraph*{Preferential Attachment (PA) similarity} We define three metrics:
\begin{itemize}
\item Preferential Attachment similarity (PA):
\begin{equation}
s_{ij}^{PA}= \sum_{\alpha}R_{i\alpha}\sum_{\alpha}R_{j\alpha}.
\end{equation}
\item Temporal Preferential Attachment similarity with power-law time-lag decay (TPA):
\begin{equation}
s_{ij}^{TPA}= \sum_{\alpha}R_{i\alpha}\sum_{\alpha}R_{j\alpha}|t_{i\alpha}-t_{j\alpha}|^{-1}(1-\delta_{t_{i\alpha},t_{j\alpha}}).
\end{equation}
\item Temporal Preferential Attachment similarity with one-step time-lag decay (TPA1):
\begin{equation}
s_{ij}^{TPA1}= \sum_{\alpha}R_{i\alpha}\sum_{\alpha}R_{j\alpha}\delta_{|t_{i\alpha}-t_{j\alpha}|,1}.
\end{equation}
\end{itemize}

In all the temporal similarity methods above, we set $(t_{i\alpha}-t_{j\alpha})^{-1}=0$ when $t_{i\alpha}=t_{j\alpha}$. Note that in the TC metrics, the factor $1-\delta_{t_{i\alpha},t_{j\alpha}}$ makes sure that events where $t_{i\alpha}=t_{j\alpha}$ do not contribute to the similarity. Indeed, when $t_{i\alpha}=t_{j\alpha}$, $i$ is not the node that infected $j$; therefore, $i$ and $j$ are unlikely to be connected in the networks. Note that in other problems such as link prediction and recommendation, the case $t_{i\alpha}=t_{j\alpha}$ may need to be treated differently.

\subsection{SIR spreading dynamics}

In the SIR model, each node is in one of the three states: Susceptible (S), Infected (I), Recovered (R).
Each node has a probability $f$ to be an initiator of the spreading process; therefore, there are $f\times N$ simultaneous initiators, on average, for each spreading process.
At each time step, each infected node can infect each of its neighbors with probability $\beta$; each infected node can recover with probability $\mu$. For simplicity, we fix $\mu=1$ (each node recovers one step after having been infected).
The process ends when there are no more infected nodes in the system.
For each empirical network, we run $50$ independent realizations of the SIR dynamics. For each process $\alpha$, we record the temporal list of the nodes infected by that process. The bipartite adjacency matrix $\mathsf{R}$ records which nodes were infected by which process: $R_{i\alpha}=1$ if $i$ has been infected by $\alpha$, whereas $R_{i\alpha}=0$ otherwise. If $R_{i\alpha}=1$, the time step at which $i$ was infected by $\alpha$ is recorded in $t_{i\alpha}$.

\subsection{Generation of the synthetic networks}
We use two well-known models for the generation of synthetic networks: the Barab\'asi-Albert (BA) model~\citep{barabasi1999emergence}, and the Small-World (SW) model~\citep{watts1998collective}.

\paragraph{Barab\'asi-Albert (BA)}

We generate networks composed of $N=500$ nodes. Our initial condition is a regular network where each node composed of $m_0=9$ nodes;
each initial node has the degree equal to $\braket{k}=5$. At each time step $t$, we add a new node to the network. The new node connects with $\braket{k}$ preexisting nodes; the probability that a preexisting node $i$ is selected is proportional to its degree $k_i(t)$ at time $t$.

\paragraph{Small-World (SW)}

We start from a regular ring lattice composed of $N=500$ nodes and degree $k=\braket{k}=5$: we connect each of the $N$ nodes with its nearest $k$ neighbors. We rewire each link with probability $p$ -- in this work, we set $p=0.1$. More specifically, for each node $i$, we select a node $j$ from its neighbors and we extract a random number $r$ from the uniform distribution in $(0,1)$. If $p$ is larger than $r$, we and remove the edge between node $i$ and node $j$, we randomly select a node $m$, and we establish an edge between node $i$ and node $m$.

\section*{Competing interests}

The authors declare that they have no competing interests.

\section*{Author's contribution}
The work presented in this paper corresponds to a collaborative development by all authors. Conceptualization, H.L., M.S.M, and M-Y.Z.; Data Curation, M-K.L. and H.L.; Formal Analysis, H.L., M-K.L. and M.S.M.; Funding~Acquisition, H.L. and M-Y.Z.; Methodology, M.S.M.; Resources, H.L. and M-Y.Z.; Software, M-K.L. and X-T.W.; Writing---Original Draft, M.S.M., M-K.L., M-Y.Z., X-T.W. and H.L.

\section*{Acknowledgements}
We wish to thank Prof. Ginestra Bianconi and Prof. Chi Ho Yeung for providing us valuable suggestions.
H.L and M.Y.Z acknowledge financial support from the National Natural Science Foundation of China (Grant Nos. 61803266, 61703281), Guangdong Province Natural Science Foundation (Grant Nos. 2016A030310051,2017A030310374,
2017B030314073), Guangdong Pre-national project (Grant Nos. 2014GK\\XM054), Shenzhen Fundamental Research Foundation ( JCYJ20160520162743717,  JCYJ20150529164-656096), Natural Science Foundation of SZU (Grant No. 2016-24), Foundation for Distinguished Young Talents in Higher Education of Guangdong, China(Grant No. 2015K-QNCX143). MSM acknowledges the University of Zurich for support through the URPP Social Networks.

\section*{Supplementary}

\subsection*{\centering Data Description}

Here we describe the $40$ empirical networks analyzed in the main text. \\

\begin{itemize}
\item[1)]

\textbf{Facebook}: a social network which contains Facebook user--user friendships.~\citep{konect:McAuley2012}
\item[2)]
\textbf{Jazz}: a music collaboration network obtained from the Red Hot Jazz Archive digital database. It includes 198 bands that performed between 1912 and 1940, with most of the bands from 1920 to 1940.~\citep{konect:arenas-jazz}
\item[3)]
\textbf{Residence hall}: a network which contains friendship ratings between 217 residents living at a residence hall located in the Australian National University campus.~\citep{konect:freeman1998}
\item[4)]
\textbf{E.coli}: a metabolic network of E.coli.~\citep{jeong2000large}
\item[5)]
\textbf{Physicians}: a network which captures the spreading paths of an innovation among 246 physicians in for towns in Illinois, Peoria, Bloomington, Quincy and Galesburg.~\citep{konect:coleman1957}
\item[6)]
\textbf{Neural}: a neural network in C. elegans.~\citep{duch2005community}
\item[7)]
\textbf{Usair}: the US air transportation network that connects airport located in the United States.
\item[8)]
\textbf{Dublin}: a network which describes the face-to-face behavior of people during the exhibition "infectious: stay away" in 2009 at the Science Gallery in Dublin.~\citep{konect:sociopatterns}
\item[9)]
\textbf{Crim}: a network which connects persons who appeared in at least one crime case as either a suspect, a victim, a witness or both a suspect and victim at the same time.
\item[10)]
\textbf{Caenorhabditis elegans}: a metabolic network of the roundworm Caenorhabditis elegans.~\citep{konect:duch05}
\item[11)]
\textbf{Email}: an email communication network at the University Rovira i Virgili in Tarragona in the south of Catalonia in Spain~\citep{konect:guimera03}
\item[12)]
\textbf{Blogs}: a network which contains front-page hyperlinks between blogs in the context of the 2004 US election.~\citep{konect:adamic2005}
\item[13)]
\textbf{Air traffic control}: a network which was constructed from the USA's FAA (Federal Aviation Administration) National Flight Data Center (NFDC), Preferred Routes Database.
\item[14)]
\textbf{Human protein}: a network of interactions between proteins in Humans (Homo sapiens), from the first large-scale study of protein–protein interactions in Human cells using a mass spectrometry-based approach.~\citep{konect:stelzl}
\item[15)]
\textbf{Hamsterster friendships}: a network which contains friendships between users of the website hamsterster.com.
\item[16)]
\textbf{UC Irvine messages}: a network which contains sent messages between the users of an online community of students from the University of California, Irvine.~\citep{konect:opsahl09}
\item[17)]
\textbf{Adolescent health}: a network which was created from a survey that took place in 1994/1995.~\citep{konect:moody}
\item[18)]
\textbf{Advogato}: a network from an online community platform for developers of free software launched in 1999.~\citep{konect:massa09}
\item[19)]
\textbf{Euroroad}: an international E-road network, a road network located mostly in Europe.~\citep{konect:eroads}
\item[20)]
\textbf{Highschool}: a network which contains friendships between boys in a small high school in Illinois.~\citep{konect:coleman}
\item[21)]
\textbf{Hypertext}: a network of face-to-face contacts between the attendees of the ACM Hypertext 2009 conference.~\citep{konect:sociopatterns}
\item[22)]
\textbf{IUI}: a network of the collaborations among the authors of papers published in Informatica and Uporabna informatika.~\citep{gao2015multi}
\item[23)]
\textbf{Amazon}: a network between web pages in amazon.com.~\citep{vsubelj2012ubiquitousness}
\item[24)]
\textbf{SCSC}: a network of collaborations between Slovenian computer scientists.~\citep{blagus2012self}
\item[25)]
\textbf{Zachary karate club}: the well-known Zachary karate club social network.~\citep{konect:ucidata-zachary}
\item[26)]
\textbf{Polbooks}: a network of books about US politics published around the time of the 2004 presidential election and sold by the online bookseller Amazon.com.~\citep{al2011survey}
\item[27)]
\textbf{Powergrid}: the power grid of the Western States of the United States of America.~\citep{konect:duncan98}
\item[28)]
\textbf{Subelj}: the software class dependency network of the JUNG 2.0.1 and javax 1.6.0.7 libraries, namespaces edu.uci.ics.jung and java/javax.
\item[29)]
\textbf{PPI}: a protein-protein interaction network.~\citep{von2002comparative}
\item[30)]
\textbf{Openflights}: a network which contains flights between airports of the world.~\citep{konect:opsahl2010b}
\item[31)]
\textbf{Bible}: a network which contains nouns (places and names) of the King James Version of the Bible and information about their co-occurrences.~\citep{zhang2013potential}
\item[32)]
\textbf{Chicago}: a network on the road transportation of the Chicago region (USA).~\citep{konect:tntp-chicago1,konect:tntp-chicago2}
\item[33)]
\textbf{DNC email}: a network of emails in the 2016 Democratic National Committee email leak.~\citep{yu2015friend}
\item[34)]
\textbf{Word}: an adjacency network of common adjectives and nouns in the novel David Copperfield written by Charles Dickens.~\citep{konect:adjnoun}
\item[35)]
\textbf{Football}: a network of American football games between Division IA colleges during regular season Fall 2000.~\citep{girvan2002network,tribuvson2016identifying}
\item[36)]
\textbf{Little Rock Lake}: the food web of Little Rock Lake, Wisconsin in the United States of America.~\citep{konect:little-rock-lake}
\item[37)]
\textbf{Unicode}: a bipartite network denotes which languages are spoken in which countries. Here we transferred it to a unipartite network.~\citep{wang2017bias}
\item[38)]
\textbf{Netsci}: a coauthorship network between scientists who published on the topic of network science.~\citep{konect:adjnoun}
\item[39)]
\textbf{TAP}: a yeast protein binding network generated by tandem affinity purification experiments.~\citep{gavin2006proteome}
\item[40)]
\textbf{Slavko}: a small friendship network from an online website.~\citep{blagus2012self}
\end{itemize}

\newpage
\subsection*{\centering SI Tables}

\begin{table*}[h!]
  \centering
   \caption{Properties of the analyzed empirical networks. $N$ is the number of nodes. $E$ is the number of edges. The parameter $\langle k\rangle$, refers to the average node degree. $C$ is the average clustering coefficient. $\beta_c = \langle k\rangle/(\langle k^2\rangle - \langle k\rangle)$ represents the critical value of the transmission probability in the SIR model in the mean-field scenario~\citep{pastor2015epidemic}.} \label{tab1}
\setlength{\tabcolsep}{5mm}{\begin{tabular}{l|c|c|c|c|c|c}
\hline
Network & $N$ & $E$ & $\langle k\rangle$ & $C$ & $\beta_c$ & url\\
\hline
Zachary karate club (Zkc) & 34 & 78 & 4.58 & 0.12 & 0.1688 & \href{http://konect.uni-koblenz.de/networks/ucidata-zachary}{url}\\
Highschool (Highs) & 70 & 366 & 10.45 & 0.29 & 0.1487 & \href{http://konect.uni-koblenz.de/networks/moreno_highschool}{url}\\
Polbooks (Polbs) & 105 & 441 & 8.4 & 0.15 & 0.2067 & \href{http://www-personal.umich.edu/~mejn/netdata/}{url}\\
Word & 112 & 425 & 3.79 & 0.17 & 0.0783 & \href{http://konect.uni-koblenz.de/networks/adjnoun_adjacency}{url}\\
Hypertext (Hypert) & 113 & 20818 & 368.46 & 0.26 & 0.0392 & \href{http://konect.uni-koblenz.de/networks/sociopatterns-hypertext}{url}\\
Football (Footb) & 115 & 1231 & 21.4 & 0.17 & 0.1623 & \href{https://icon.colorado.edu/#!/networks}{url}\\
Little Rock Lake (LRL) & 183 & 2494 & 27.25 & 0.09 & 0.0229 & \href{http://konect.uni-koblenz.de/networks/maayan-foodweb}{url}\\
Jazz & 198 & 2742 & 27.69 & 0.62 & 0.0266 & \href{http://konect.uni-koblenz.de/networks/arenas-jazz}{url}\\
Residence hall (Rhall) & 217 & 2672 & 24.62 & 0.24 & 0.0688 & \href{http://konect.uni-koblenz.de/networks/moreno_oz}{url}\\
E.coli & 230 & 695 & 6.04 & 0.22 & 0.0752 & \href{http://www.weizmann.ac.il/mcb/UriAlon/e-coli-transcription-network}{url}\\
Physicians (Phys) & 241 & 1098 & 9.11 & 0.13 & 0.1366 & \href{http://konect.uni-koblenz.de/networks/moreno_innovation}{url}\\
Neural & 297 & 2359 & 15.88 & 0.12 & 0.049 & \href{https://icon.colorado.edu/#!/networks}{url}\\
USAir & 332 & 2126 & 12.8 & 0.63 & 0.0231 & \href{http://vlado.fmf.uni-lj.si/pub/networks/data/default.htm}{url}\\
Slavko & 334 & 2218 & 13.28 & 0.17 & 0.0791 & \href{http://wwwlovre.appspot.com/support.jsp}{url}\\
Netsci & 379 & 914 & 4.82 & 0.74 & 0.1424 & \href{https://icon.colorado.edu/#!/networks}{url}\\
Dublin & 410 & 2765 & 13.48 & 0.30 & 0.1044 & \href{http://konect.uni-koblenz.de/networks/sociopatterns-infectious}{url}\\
Caenorhabditis elegans (Cae) & 453 & 4596 & 10.15 & 0.07 & 0.0465 & \href{http://konect.uni-koblenz.de/networks/arenas-meta}{url}\\
Unicode (Unic) & 767 & 1255 & 3.27 & 0.01 & 0.0455 & \href{http://konect.uni-koblenz.de/networks/unicodelang}{url}\\
Scsc & 961 & 1925 & 4.01 & 0.02 & 0.2033 & \href{http://wwwlovre.appspot.com/support.jsp}{url}\\
Email & 1133 & 5451 & 9.62 & 0.22 & 0.0565 & \href{http://konect.uni-koblenz.de/networks/arenas-email}{url}\\
Euroroad (Eroad) & 1174 & 1417 & 2.41 & 0.01 & 0.1563 & \href{http://konect.uni-koblenz.de/networks/subelj_euroroad}{url}\\
Blogs & 1224 & 19025 & 31.08 & 0.14 & 0.0123 & \href{http://konect.uni-koblenz.de/networks/moreno_blogs}{url}\\
Air traffic control (Air.tra) & 1226 & 2615 & 4.26 & 0.02 & 0.2353 & \href{http://konect.uni-koblenz.de/networks/maayan-faa}{url}\\
TAP & 1373 & 6833 & 9.96 & 0.53 & 0.0651 & \href{https://www3.nd.edu/~networks/resources.htm}{url}\\
Crim & 1380 & 1476 & 2.13 & 0.1 & 0.0458 & \href{http://konect.uni-koblenz.de/networks/moreno_crime}{url}\\
Chicago (Chic) & 1467 & 1298 & 1.76 & 0 & 0.1411 & \href{http://konect.uni-koblenz.de/networks/tntp-ChicagoRegional}{url}\\
Human protein (HP) & 1706 & 6207 & 7.27 & 0.02 & 0.0653 & \href{http://konect.uni-koblenz.de/networks/maayan-Stelzl}{url}\\
Bible & 1773 & 16401 & 18.5 & 0.1 & 0.1299 & \href{http://konect.uni-koblenz.de/networks/moreno_names}{url}\\
Hamsterster friendships (HF) & 1858 & 12534 & 13.49 & 0 & 0.0217 & \href{http://konect.uni-koblenz.de/networks/petster-friendships-hamster}{url}\\
UC Irvine messages (UC.irv) & 1899 & 59835 & 63.01 & 0.04 & 0.0317 & \href{http://konect.uni-koblenz.de/networks/opsahl-ucsocial}{url}\\
DNC emails (DNC) & 2029 & 39264 & 38.70 & 0.06 & 0.0164 & \href{http://konect.uni-koblenz.de/networks/dnc-temporalGraph}{url}\\
IUI & 2288 & 4190 & 3.66 & 0.03 & 0.3068 &\href{http://wwwlovre.appspot.com/support.jsp}{url}\\
PPI & 2375 & 11693 & 9.84 & 0.3 & 0.0301 & \href{http://interactome.dfci.harvard.edu/A_thaliana/index.php?page=download}{url}\\
Adolescent health (Health) & 2539 & 12969 & 10.21 & 0.10 & 0.1408 & \href{http://konect.uni-koblenz.de/networks/moreno_health}{url}\\
Amazon (Ama) & 2880 & 5037 & 3.49 & 0.01 & 0.1651 & \href{http://wwwlovre.appspot.com/support.jsp}{url}\\
Facebook (Faceb) & 2888 & 2981 & 2.06 & 0 & 0.1879 & \href{http://konect.uni-koblenz.de/networks/ego-facebook}{url}\\
Openflights (Oflgs) & 2939 & 30501 & 20.75 & 0.39 & 0.0184 & \href{http://konect.uni-koblenz.de/networks/opsahl-openflights}{url}\\
Powergrid (Pgrid) & 4941 & 6594 & 2.66 & 0.01 & 0.1175 & \href{http://konect.uni-koblenz.de/networks/opsahl-powergrid}{url}\\
Subelj & 6434 & 150985 & 46.93 & 0.09 & 0.0513 & \href{http://konect.uni-koblenz.de/networks/subelj_jdk}{url}\\
Advogato (Adv) & 6541 & 51127 & 7.82 & 0.11 & 0.0171 & \href{http://konect.uni-koblenz.de/networks/advogato}{url}\\
\hline

\end{tabular}}
\end{table*}
\pagebreak
\newpage
\begin{table*}[h]
  \centering
  \caption{Basic properties of real networks and the performance of the COS, TCOS, TCOS1, SS, TSS, TSS1, LHN, TLHN, TLHN1, HD, THD and THD1 methods on these networks, based on the AUC metric. The parameters are set as $\beta$ = 4 $\beta_c$ and $f=0.5$. We select a relatively large $\beta$ because the performance difference between traditional similarity metric and temporal similarity metric becomes more significant under large $\beta$. The similarity method with the best performance in each network is highlighted in bold font. The results are averaged over $50$ independent realizations.}
 \resizebox{\textwidth}{65mm}{\begin{tabular}
{l|cc|ccc|ccc|ccc|ccc}
\hline
\multirow{2}{*}{Network} &
\multicolumn{2}{c|}{Basic properties} &                                                                         \multicolumn{12}{c}{AUC} \\

\cline{2-15}
 &N &E  &COS	 &	TCOS	 &	TCOS1 &	SSI	&	TSSI	 &	TSSI1	 &	LHN	 &	TLHN	 &	TLHN1	 &	HDI	 &	THDI	 &	THDI1\\	
 \hline
Zkc & 34 & 78 & 0.6683 & \textbf{0.6816} & 0.6811&0.6659&0.6797&\textbf{0.6835}&0.6585&0.6770&\textbf{0.6776}&0.6604&\textbf{0.6782}&0.6775\\
Highs& 70 & 366& 0.7614&\textbf{0.8532}&0.8525&0.7600&\textbf{0.8491}&0.8490&0.6396&\textbf{0.8367}&0.8367&0.7500&\textbf{0.8411}&0.8410\\
Polbs & 105 & 441 & 0.7201&\textbf{0.7206}&0.7199&0.7073&0.7096&\textbf{0.7101}&0.6552&0.6711&\textbf{0.6714}&0.6888&0.7002&\textbf{0.7011}\\

  Word     &	112	 &	425	  &	0.8413	& 	0.9350	&	\textbf{0.9506}	&	0.7821 	 &	0.8898	&	\textbf{0.9305}	&	0.8311	&	0.9033	 &	\textbf{0.9396}	&	0.8056	&	0.8937	&	\textbf{0.9313}	 \\
  Hypert & 113 & 20818 &0.5314&\textbf{0.5447}&0.5438&0.5219&\textbf{0.5356}&0.5352&0.5446&0.5491&\textbf{0.5509}&0.5137&\textbf{0.5271}&0.5271\\
  Footb & 115&1231&0.6789&\textbf{0.7216}&0.7216&0.6557&\textbf{0.6945}&0.6941&0.6813&\textbf{0.7116}&0.7108&0.6373&\textbf{0.6750}&0.6741\\
  LRL   &	183	 &	2494 	 &	0.6462	&	0.6488	&	\textbf{0.6491}	&	0.6452 	 &	\textbf{0.6482}	&	0.6481	&	0.6431	&	0.6469	 &	\textbf{0.6469}	&	0.6464	&	0.6488	&	\textbf{0.6488}	\\
  Jazz     &	198	 &	2742 	 &	0.8312	&	0.8650	&	\textbf{0.8783}	& 0.8211	 &	0.8609	&	\textbf{0.8753}	&		0.7563	&	0.7875	&	\textbf{0.8218} & 0.7942	&	0.8487	 &	\textbf{0.8686}		 \\
  Rhall   &	217	 &	2672 	 &	0.7103	&	\textbf{0.8683}	&	0.8682	&	0.6856 	 &	0.7954	& \textbf{0.7956}	&	0.7094	&	0.8308	 &	\textbf{0.8311}	&	0.7030	&	0.8344	&	\textbf{0.8346}	\\
  E. coli   &	230	 &	695	  &	0.9018	&	0.9726	&	\textbf{0.9822}	&	0.8876 	 &	0.9659	&	\textbf{0.9797}	&	0.7924	&	0.8999	 &	\textbf{0.9307}	&	0.8965	&	0.9504	& \textbf{0.9712}	\\
  Phys   &	241	 &	1098 	 &	0.8610	&	0.8737	&	\textbf{0.8738}	&	0.8587 	 &	0.8717	&	\textbf{0.8720}	&	0.6278	&	0.8375	 &	\textbf{0.8377}	&	0.8521	&	0.8672	&	\textbf{0.8675}	\\
Neural	 &	297	 &	2359 	 &	0.7314	 &	0.8266	 &	\textbf{0.8380}	 &	0.7287 	 &	0.8098	 &	\textbf{0.8303}	 &	0.7174	 &	0.7865	 &	\textbf{0.8068}	 &	0.7213	 &	0.7920	 &	\textbf{0.8176}	 \\
  USAir   &	332	 &	2126	 &	0.9135	&	0.9435	&	\textbf{0.9539}	&	0.9089	  &	0.9365	&	\textbf{0.9474}	&	0.7852	&	0.8199	 &	\textbf{0.8273}	&	0.9023	&	0.9289	&	\textbf{0.9390}	\\
  Slavko & 334 & 2218 &\textbf{0.7711}&0.7686&0.7683&0.7596&\textbf{0.7615}&0.7609&0.7552&\textbf{0.7576}&0.7574&0.7508&\textbf{0.7549}&0.7549\\
  Netsci &379&914&0.9225&\textbf{0.9755}&0.9752&0.9198&\textbf{0.9760}&0.9760&0.9138&\textbf{0.9635}&0.9632&0.8987&\textbf{0.9810}&0.9807\\
  Dublin   &	410	 &	2765 	 &	0.7959	&	0.8460	&	\textbf{0.8462}	&	0.7322 	 &	0.8427	&	\textbf{0.8430}	&	0.6985	&	0.7567	 &	\textbf{0.7567}	&	0.7735	&	0.8226	&	\textbf{0.8231}	\\
 Cae    &    453  & 4596 & 0.7122 & \textbf{0.7369} & 0.7348 & 0.7062 & 0.7202 & \textbf{0.7231} & 0.6794 & 0.7161 & \textbf{0.7190} & 0.6992 & 0.7160 & \textbf{0.7163}\\
 Unic &767&1255&0.6018&\textbf{0.6020}&0.6020&0.6019&0.6021&\textbf{0.6023}&\textbf{0.6022}&0.6021&0.6020&\textbf{0.6021}&0.6021&0.6020\\
 Scsc & 961 & 1925&0.8158&0.8160&\textbf{0.8161}&0.8151&\textbf{0.8159}&0.8159&0.8158&0.8159&\textbf{0.8160}&0.8151&\textbf{0.8162}&0.8154\\
  Email   &	1133	 &	5451 	 &	0.8567	&	0.9725	&	\textbf{0.9946}	&	0.8423	 &	0.9617	&	\textbf{0.9925}	&	0.8341	&	0.9407	 &	\textbf{0.9823}	&	0.8389	&	0.9470	&	\textbf{0.9797}	 \\
  Eroad&1174&1417&0.8942&\textbf{0.9186}&0.9185&0.8943&0.9146&\textbf{0.9148}&0.8114&\textbf{0.9177}&0.9172&0.8925&\textbf{0.9136}&0.9136\\
  Blogs   &	1224	 &	19025 	 &	0.8431	&	0.8580	&	\textbf{0.8581}	&	0.8401 	 &	0.8493	&	\textbf{0.8494}	&	0.8342	&	0.8451	 &	\textbf{0.8452}	&	0.8269	&	\textbf{0.8391}	&	0.8390	\\
  Air.tra   &	1226	 &	2615 	 &	0.8439	&	0.8640	&	\textbf{0.8643}	&	0.8257 	 &	0.8533	&	\textbf{0.8537}	&	0.8330	&	0.8568	 &	\textbf{0.8571}	&	0.8527	&	0.8652	&	\textbf{0.8652}	\\
  TAP   &	1373	 &	6833 	 &	0.8964	&	0.9924	&	\textbf{0.9983}	&	0.8757 	 &	0.9906	&	\textbf{0.9982}	&	0.8423	&	0.9902	 &	\textbf{0.9980}	&	0.8558	&	0.9814	&	\textbf{0.9966}	\\
      Crim   &	1380	 &	1476 	 &	0.7959	&	0.8444	&	\textbf{0.8446}	&	0.6986 	 &	0.7556	&	\textbf{0.7558}	&	0.7870	&	0.8297	 &	\textbf{0.8298}	&	0.7738	&	0.8219	&	\textbf{0.8220}	\\
Chic & 1467 & 1298 &0.6513&0.6515&0.6517&0.6517&0.6516&0.6516&0.6517&0.6519&0.6518&0.6519&0.6517&0.6518\\
  HP   &	1706	 &	6207 	 &	0.9376	&	\textbf{0.9833}	&	0.9831	&	0.8802 	 &	0.9371	&	\textbf{0.9373}	&	0.9274	&	0.9756	 &	\textbf{0.9760}	&	0.9081	&	0.9720	&	\textbf{0.9723}	\\
  Bible&1773&16401
  &0.7413&0.7551&\textbf{0.7555}
  &0.7131&\textbf{0.7253}&0.7245
  &0.7056&0.7303&\textbf{0.7310}
  &0.6822&0.7005&\textbf{0.7007}\\
  HF   &	1858	 &	12534 	 &	0.7726	&	0.7768	&	\textbf{0.7770}	&	0.7673 	 &	0.7714	&	\textbf{0.7717}	&	0.7708	&	0.7737	 &	\textbf{0.7738}	&	0.7701	&	\textbf{0.7737}	&	0.7733	\\
  Uc.irv   &	1899	 &	59835 	 &	0.8996	&	0.9471	&	\textbf{0.9471}	&	0.8973 	 &	0.9254	&	\textbf{0.9258}	&	0.8952	&	0.9265	 &	\textbf{0.9266}	&	0.8956	&	0.9123	&	\textbf{0.9127}	\\
  DNC&2029&39264
  &0.9613&0.9618&\textbf{0.9619}
  &0.9608&0.9613&\textbf{0.9614}
  &0.9372&0.9432&\textbf{0.9437}
  &0.9603&\textbf{0.9612}&0.9611\\
  IUI&2288&4190
  &0.8276&\textbf{0.8277}&0.8274
  &\textbf{0.8281}&0.8268&0.8278
  &\textbf{0.8287}&0.8278&0.8283
  &0.8275&0.8271&\textbf{0.8278}\\
  PPI&2375&11693
  &0.9349&\textbf{0.9747}&0.9747
  &0.9336&\textbf{0.9707}&0.9706
  &0.7811&\textbf{0.9250}&0.9249
  &0.9314&\textbf{0.9652}&0.9650\\
  Health & 2539 & 12969 & 0.7561 & 0.8970 & \textbf{0.8979} & 0.7502 & 0.8886 & \textbf{0.8890} & 0.7896 & 0.9339 & \textbf{0.9347} & 0.7444 & 0.8724 & \textbf{0.8733}\\
  Ama&2880&5037
  &0.6535&0.9172&\textbf{0.9178}
  &0.6502&\textbf{0.9172}&0.9172
  &0.6531&\textbf{0.9168}&0.9160
  &0.6314&0.9138&\textbf{0.9140}\\
  Faceb&2888&2981
  &0.6789&\textbf{0.7216}&0.7216
  &0.6557&\textbf{0.6945}&0.6941
  &0.6813&\textbf{0.7116}&0.7108
  &0.6373&\textbf{0.6750}&0.6741\\
  Oflgs&2939&30501
  &0.9384&\textbf{0.9581}&0.9578
  &0.9311&0.9505&\textbf{0.9507}
  &0.6979&\textbf{0.7957}&0.7957
  &0.9239&0.9432&\textbf{0.9433}\\
  Pgrid&4941&6594
  &0.8927&0.8928&\textbf{0.8929}
  &0.8922&\textbf{0.8929}&0.8926
  &0.8916&\textbf{0.8928}&0.8927
  &0.8917&\textbf{0.8928}&0.8927\\
  Subelj&6434&150985
  &0.6284&\textbf{0.7016}&0.7016
  &0.6353&\textbf{0.6745}&0.6741
  &0.6616&\textbf{0.6916}&0.6908
  &0.6178&\textbf{0.6550}&0.6541\\
  Adv & 6541 & 51127 & 0.9104 & 0.9269 & \textbf{0.9270} & 0.9099 & 0.9242 & \textbf{0.9249} & 0.7914 & \textbf{0.8373} & 0.8361 & 0.9050 & \textbf{0.9180} & 0.9173\\
\hline
    \end{tabular}}
\end{table*}
\newpage

\begin{table*}[h]
  \centering
  \caption{Basic properties of real networks and the performance of the COS, TCOS, TCOS1, SS, TSS, TSS1, LHN, TLHN, TLHN1, HD, THD and THD1 methods on these networks, based on the precision metric. The parameters are set as $\beta$ = 4 $\beta_c$ and $f=0.5$. We select a relatively large $\beta$ because the performance difference between traditional similarity metric and temporal similarity metric becomes more significant under large $\beta$. The similarity method with the best performance in each network is highlighted in bold font. The results are averaged over $50$ independent realizations.}
\resizebox{\textwidth}{65mm}{\begin{tabular}{l|cc|ccc|ccc|ccc|ccc}
\hline
\multirow{2}{*}{Network} &
\multicolumn{2}{c|}{Basic properties} &
\multicolumn{12}{c|}{Precision} \\
\cline{2-15}

 &N &E  &COS	 &	TCOS	 &	TCOS1 &	SSI	&	TSSI	 &	TSSI1	 & LHN	 &	TLHN	 &	TLHN1	 &	HDI	 &	THDI	 &	THDI1\\	
 \hline
Zkc & 34 & 78 & 0.207 & 0.235 & \textbf{0.236}&0.209&\textbf{0.237}&0.230&0.193&0.214&\textbf{0.221}&0.209&0.224&\textbf{0.230}\\
Highs& 70 & 366& 0.365&0.535&\textbf{0.610}&0.365&0.536&\textbf{0.611}&0.181&0.474&\textbf{0.571}&0.357&0.537&\textbf{0.620}\\
Polbs & 105 & 441 & 0.217&0.217&\textbf{0.241}&0.217&0.213&\textbf{0.234}&0.138&0.172&\textbf{0.186}&0.205&0.198&\textbf{0.220}\\
  Word     &	112	 &	425	  &  0.313 &	0.568	& 	\textbf{0.624}	&  0.307	&	0.548 	 &	\textbf{0.610}	&	0.241	&	0.320	&	\textbf{0.390}	 &	0.297	&	0.489	&	\textbf{0.562}	 \\
    Hypert & 113 & 20818 &0.221&0.239&\textbf{0.244}&0.215&0.231&\textbf{0.240}&0.220&0.231&\textbf{0.233}&0.202&0.221&\textbf{0.228}\\
      Footb & 115&1231&0.224&0.286&\textbf{0.320}&0.212&0.267&\textbf{0.306}&0.156&0.201&\textbf{0.232}&0.176&0.233&\textbf{0.278}\\
  LRL   &	183	 &	2494 	 &	0.282	&	\textbf{0.310}	&	0.287	&	0.275 	 &	\textbf{0.305}	&	0.286	&	0.259	&	\textbf{0.294}	 &	0.287	&	0.270	&	\textbf{0.302}	&	0.285	\\
  Jazz     &	198	 &	2742 	 &	0.434	&	0.531	&	\textbf{0.538}	&	0.426	 &	0.528	&	\textbf{0.537}	&	0.304	&	0.342	 &	\textbf{0.407}	&	0.405	&	0.509	&	\textbf{0.531}	 \\
  Rhall   &	217	 &	2672 	 &	0.189	&	0.462	&	\textbf{0.565}	&	0.181 	 &	0.340	& \textbf{0.457}	&	0.189	&	0.379	 &	\textbf{0.506}	&	0.185	&	0.408	&	\textbf{0.520}	\\

  E. coli   &	230	 &	695	  &	0.323	&	0.584	&	\textbf{0.652}	&	0.348 	 &	0.558	&	\textbf{0.626}	&	0.278	&	0.303	 &	\textbf{0.354}	&	0.317	&	0.514	&	\textbf{0.579}	\\
   Phys   &	241	 &	1098 	 &	0.315	&	0.460	&	\textbf{0.464}	&	0.162 	 &	0.275	&	\textbf{0.353}	&	0.309	&	0.439	 &	\textbf{0.468}	&	0.284	&	0.427	&	\textbf{0.453}	\\
Neural	 &	297	 &	2359 	 &	0.178	 &	0.327	 &	\textbf{0.376}	 &	0.152 	 &	0.312	 &	\textbf{0.369}	 &	0.124	 &	0.243	 &	\textbf{0.295}	 &	0.142	 &	0.271	 &	\textbf{0.335}	 \\
  USAir   &	332	 &	2126	 &	\textbf{0.561}	&	0.555	&	0.541	&	\textbf{0.554}	  &	0.549	&	0.536	&	0.219  &	0.234	&	\textbf{0.339}	 	&	0.526	&	0.532	&	\textbf{0.533}	\\
  Slavko & 334 & 2218 &0.253&\textbf{0.261}&0.258&0.243&\textbf{0.254}&0.253&0.132&0.162&\textbf{0.175}&0.215&0.229&\textbf{0.236}\\
    Netsci &379&914&0.371&0.581&\textbf{0.771}&0.370&0.583&\textbf{0.773}&0.279&0.0.517&\textbf{0.628}&0.359&0.579&\textbf{0.788}\\
  Dublin   &	410	 &	2765 	 &	0.203	&	0.355	&	\textbf{0.361}	&	0.115 	 &	0.212	&	\textbf{0.279}	&	0.199	&	0.343	 &	\textbf{0.370}	&	0.176	&	0.319	&	\textbf{0.348}	\\
 Cae    &    453  & 4596 & 0.141 & 0.229 & \textbf{0.236} & 0.139 & 0.224 & \textbf{0.233} & 0.077 & 0.101 & \textbf{0.114} & 0.112 & 0.195 & \textbf{0.214}\\
  Unic &767&1255&0.104&\textbf{0.156}&0.153&0.103&0.152&\textbf{0.153}&0.098&0.141&\textbf{0.152}&0.101&0.148&\textbf{0.152}\\
  Scsc & 961 & 1925&0.233&0.271&\textbf{0.279}&0.230&0.266&\textbf{0.272}&0.179&0.214&\textbf{0.234}&0.216&0.250&\textbf{0.258}\\
  Email   &	1133	 &	5451 	 &	0.467	&	0.568	&	\textbf{0.766}	&	0.442	 &	0.503	&	\textbf{0.707}	&	0.145	&	0.206	 &	\textbf{0.341}	&	0.256	&	0.397	&	\textbf{0.581}	 \\
 Eroad&1174&1417&0.230&0.424&\textbf{0.425}&0.227&0.424&\textbf{0.425}&0.137&0.185&\textbf{0.259}&0.228&0.413&\textbf{0.414}\\
  Blogs   &	1224	 &	19025 	 &	0.173	&	0.242	&	\textbf{0.287}	&	0.166 	 &	0.216	&	\textbf{0.246}	&	0.173	&	0.226	 &	\textbf{0.254}	&	0.171	&	0.226	&	\textbf{0.256}	\\
  Air.tra   &	1226	 &	2615 	 &	0.186	&	0.374	&	\textbf{0.402}	&	0.179 	 &	0.383	&	\textbf{0.403}	&	0.056	&	0.132	 &	\textbf{0.156}	&	0.158	&	0.366	&	\textbf{0.367}	\\
  TAP   &	1373	 &	6833 	 &	0.232	&	0.635	&	\textbf{0.734}	&	0.215 	 &	0.617	&	\textbf{0.717}	&	0.158	&	0.502	 &	\textbf{0.713}	&	0.187	&	0.590	&	\textbf{0.678}	\\
  Crim   &	1380	 &	1476 	 &	0.034	&	0.227	&	\textbf{0.228}	&	0.041 	 &	0.228	&	\textbf{0.228}	&	0.077	&	0.229	 &	\textbf{0.287}	&	0.041	&	0.228	&	\textbf{0.231}	\\
  Chic & 1467 & 1298 &0.294&0.303&\textbf{0.304}&0.298&0.303&\textbf{0.304}&0.287&0.302&\textbf{0.304}&0.298&0.303&\textbf{0.304}\\
  HP   &	1706	 &	6207 	 &	0.135	&	0.221	&	\textbf{0.312}	&	0.122 	 &	0.189	&	\textbf{0.246}	&	0.136	&	0.199	 &	\textbf{0.255}	&	0.131	&	0.198	&	\textbf{0.255}	\\
    Bible&1773&16401
  &0.152&0.208&\textbf{0.213}
  &0.148&0.207&\textbf{0.216}
  &0.043&0.052&\textbf{0.056}
  &0.119&0.184&\textbf{0.210}\\
  HF   &	1858	 &	12534 	 &	0.090	&	0.157	&	\textbf{0.195}	&	0.067 	 &	0.075	&	\textbf{0.086}	&	0.086	&	0.127	 &	\textbf{0.159}	&	0.071	&	0.122	&	\textbf{0.151}	\\
  Uc.irv   &	1899	 &	59835 	 &	0.176	&	\textbf{0.370}	&	0.354	&	0.113 	 &	0.257	&	\textbf{0.324}	&	0.171	&	\textbf{0.374}	 &	0.371	&	0.155	&	0.339	&	\textbf{0.342}	\\
    DNC&2029&39264
  &0.373&\textbf{0.376}&0.373
  &0.372&\textbf{0.375}&0.374
  &0.013&\textbf{0.019}&0.019
  &0.362&0.370&\textbf{0.373}\\
  IUI&2288&4190
  &0.140&0.196&\textbf{0.229}
  &0.143&0.196&\textbf{0.205}
  &0.351&0.303&\textbf{0.323}
  &0.142&0.183&\textbf{0.199}\\

 Health & 2539 & 12969 & 0.025 & 0.533 & \textbf{0.618} & 0.025 & 0.530 & \textbf{0.616} & 0.039 & 0.327 & \textbf{0.478} & 0.025 & 0.418 & \textbf{0.599}\\
  Ama&2880&5037
  &0.008&\textbf{0.015}&0.015
  &0.009&0.013&\textbf{0.014}
  &0.005&\textbf{0.014}&0.014
  &0.007&0.013&\textbf{0.014}\\
  Faceb&2888&2981
  &0.224&0.286&\textbf{0.320}
  &0.212&0.268&\textbf{0.306}
  &0.157&0.201&\textbf{0.232}
  &0.176&0.234&\textbf{0.278}\\
  Oflgs&2939&30501
  &0.248&0.321&\textbf{0.368}
  &0.248&0.320&\textbf{0.363}
  &0.021&0.028&\textbf{0.031}
  &0.244&0.293&\textbf{0.335}\\
  Pgrid&4941&6594
  &0.185&0.311&\textbf{0.312}
  &0.183&0.310&\textbf{0.311}
  &0.122&0.206&\textbf{0.228}
  &0.184&0.301&\textbf{0.302}\\
  Subelj&6434&150985
  &0.067&0.124&\textbf{0.130}
  &0.065&0.120&\textbf{0.125}
  &0.016&0.046&\textbf{0.051}
  &0.066&0.114&\textbf{0.115}\\
 Adv & 6541 & 51127 & 0.061 & 0.155 & \textbf{0.322} & 0.060 & 0.151 & \textbf{0.288} & 0.002 & 0.069 & \textbf{0.091} & 0.061 & 0.120 & \textbf{0.240}\\
\hline
\end{tabular}}
\end{table*}
\pagebreak
\newpage

\subsection*{\centering SI FIGURES}
\setcounter{figure}{0}
\renewcommand{\thefigure}{S\arabic{figure}}
\begin{figure*}[h!]
  \centering
  \includegraphics[width=17cm,scale=0.5]{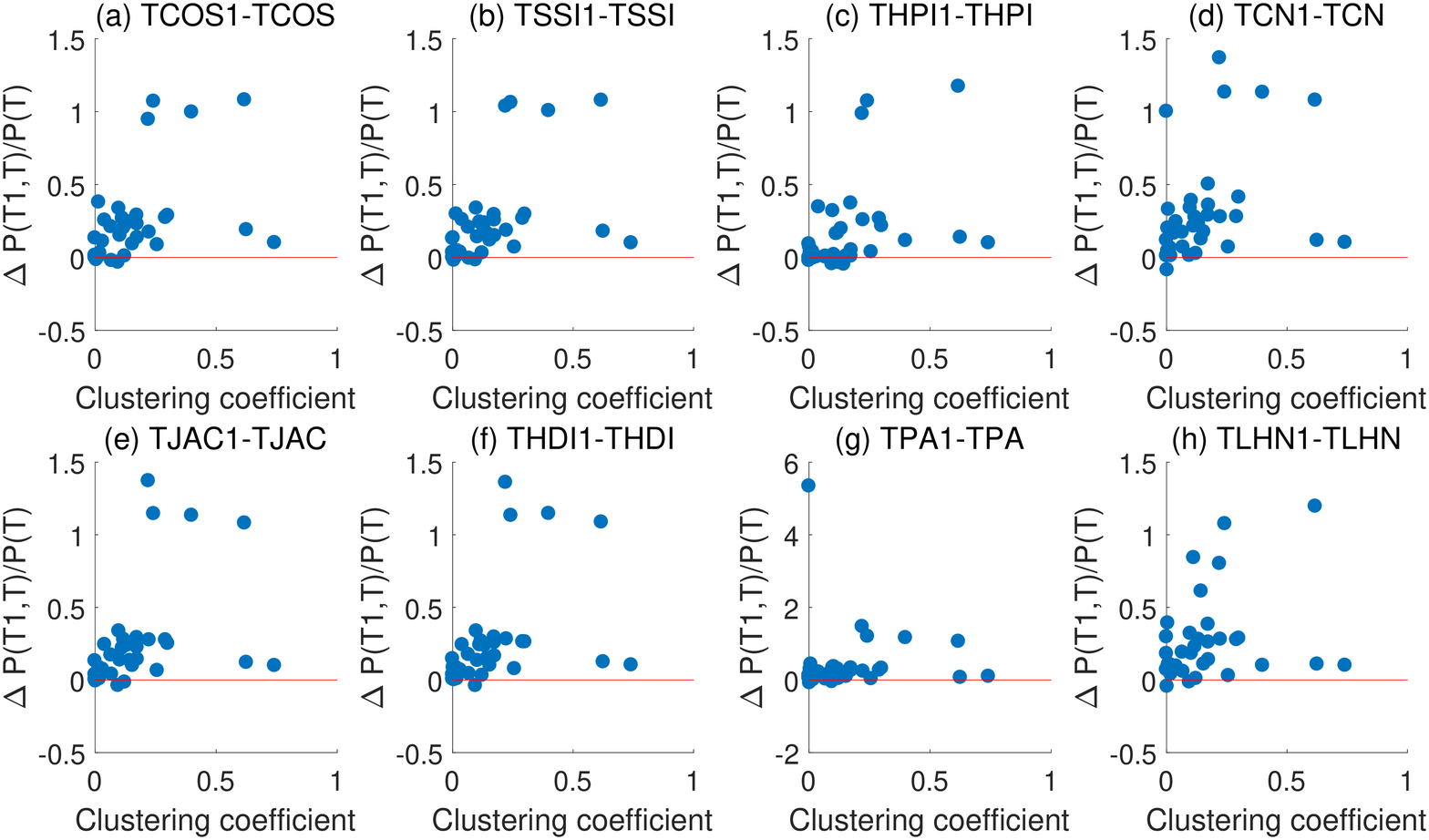}
\caption{Results for the SIR spreading dynamics: Reconstruction precision relative difference $\Delta P(T1,T)/P(T)=(P(T1)-P(T))/P(T)$ as a function of the network clustering coefficient (see Methods for the definition). Each dot represents an empirical network, we totally analyzed $40$ empirical network. We use $f = 0.5, \beta = 8 \beta_c$ here, where $\beta_c = \langle k\rangle/(\langle k^2\rangle - \langle k\rangle)$. The results are averaged over $50$ independent realizations.}
\label{SM1}
\end{figure*}

\begin{figure*}[h!]
  \centering
  \includegraphics[width=17cm,scale=0.5]{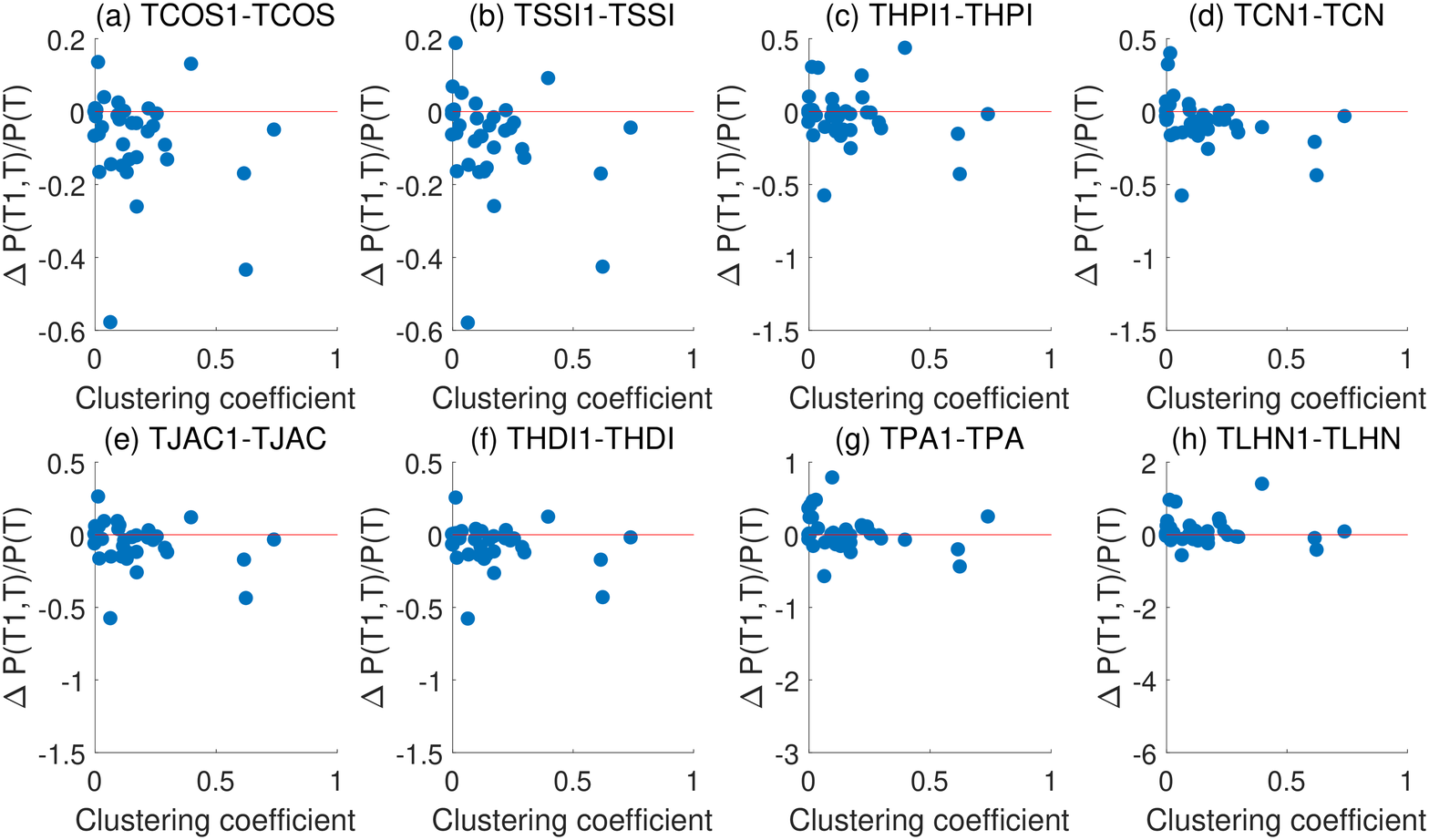}
\caption{Results for the SIR spreading dynamics: Reconstruction precision relative difference $\Delta P(T1,T)/P(T)=(P(T1)-P(T))/P(T)$ as a function of the network clustering coefficient (see Methods for the definition). Each dot represents an empirical network, we totally analyzed $40$ empirical network. We use $f = 0.5, \beta = \beta_c$ here, where $\beta_c = \langle k\rangle/(\langle k^2\rangle - \langle k\rangle)$. The results are averaged over $50$ independent realizations.}
\label{SM2}
\end{figure*}

\begin{figure*}[h!]
  \centering
  \includegraphics[width=17cm,scale=0.5]{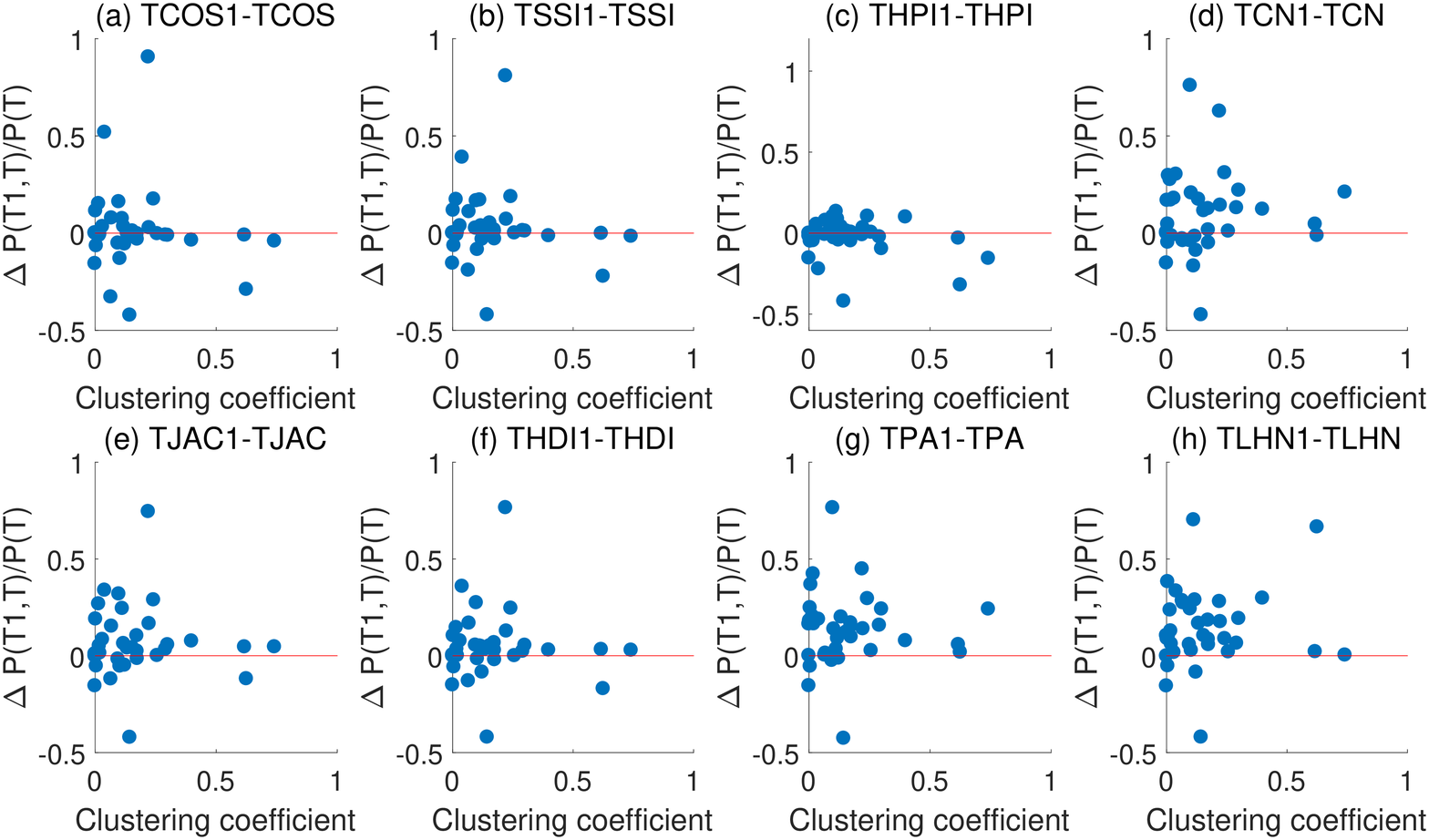}
\caption{Results for the SIR spreading dynamics: Reconstruction precision relative difference $\Delta P(T1,T)/P(T)=(P(T1)-P(T))/P(T)$ as a function of the network clustering coefficient (see Methods for the definition). Each dot represents an empirical network, we totally analyzed $40$ empirical network. We use $f = 0.5, \beta = 2\,\beta_c$ here, where $\beta_c = \langle k\rangle/(\langle k^2\rangle - \langle k\rangle)$. The results are averaged over $50$ independent realizations.}
\label{SM3}
\end{figure*}
\pagebreak
\newpage
\begin{figure*}[h!]
  \centering
  \includegraphics[width=17cm,scale=0.5]{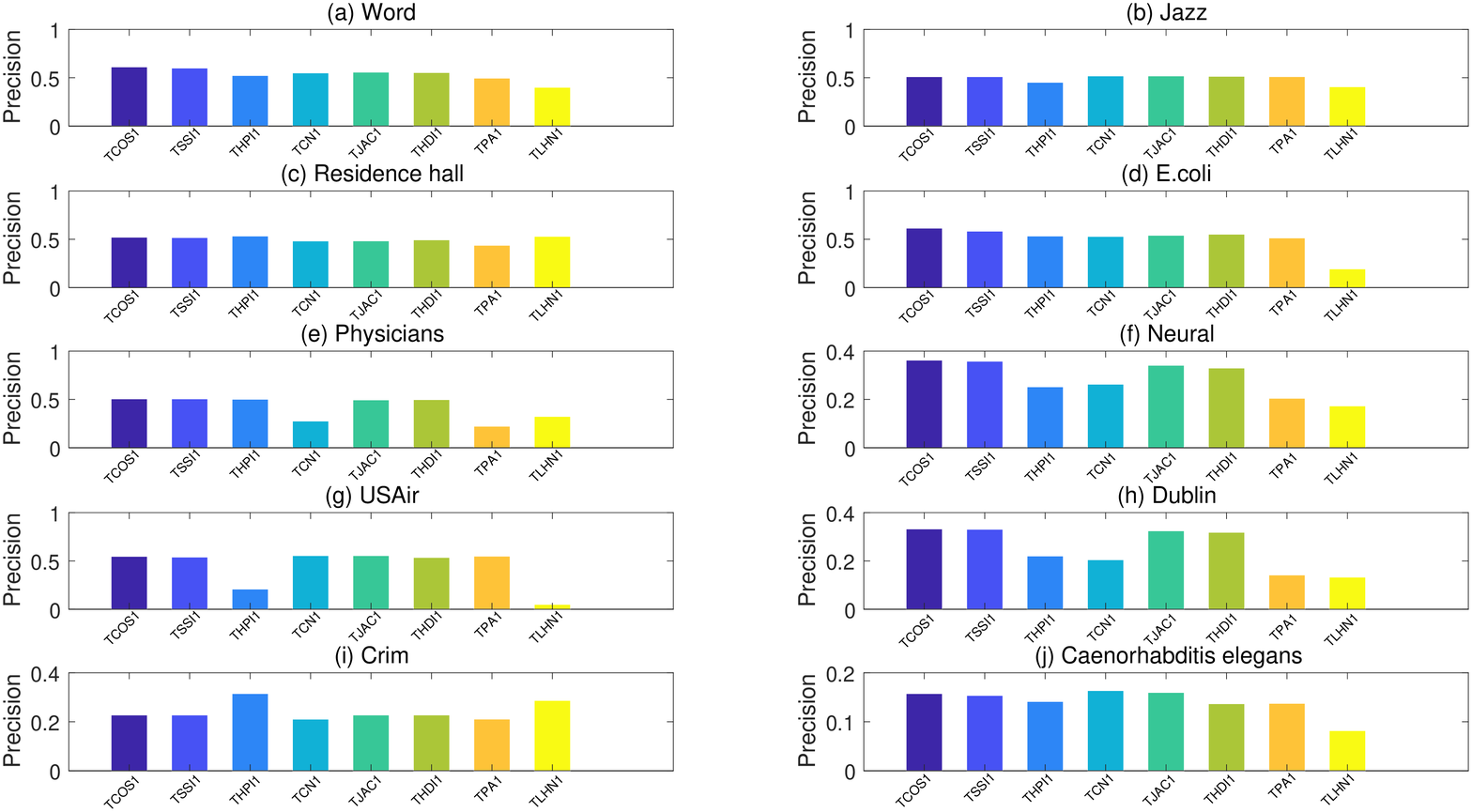}
\caption{Results for the SIR spreading dynamics: Reconstruction precision for $10$ networks. Each panel represents the performance of the $8$ TX1 metrics for a single network. This figure shows the results for $\beta = 4 \,\beta_c$, where $\beta_c = \langle k\rangle/(\langle k^2\rangle - \langle k\rangle)$, and $f = 0.5$. The results are averaged over $50$ independent realizations.}
\label{SM4}
\end{figure*}

\pagebreak
\newpage
\begin{figure*}[h!]
  \centering
  \includegraphics[width=17cm,scale=0.5]{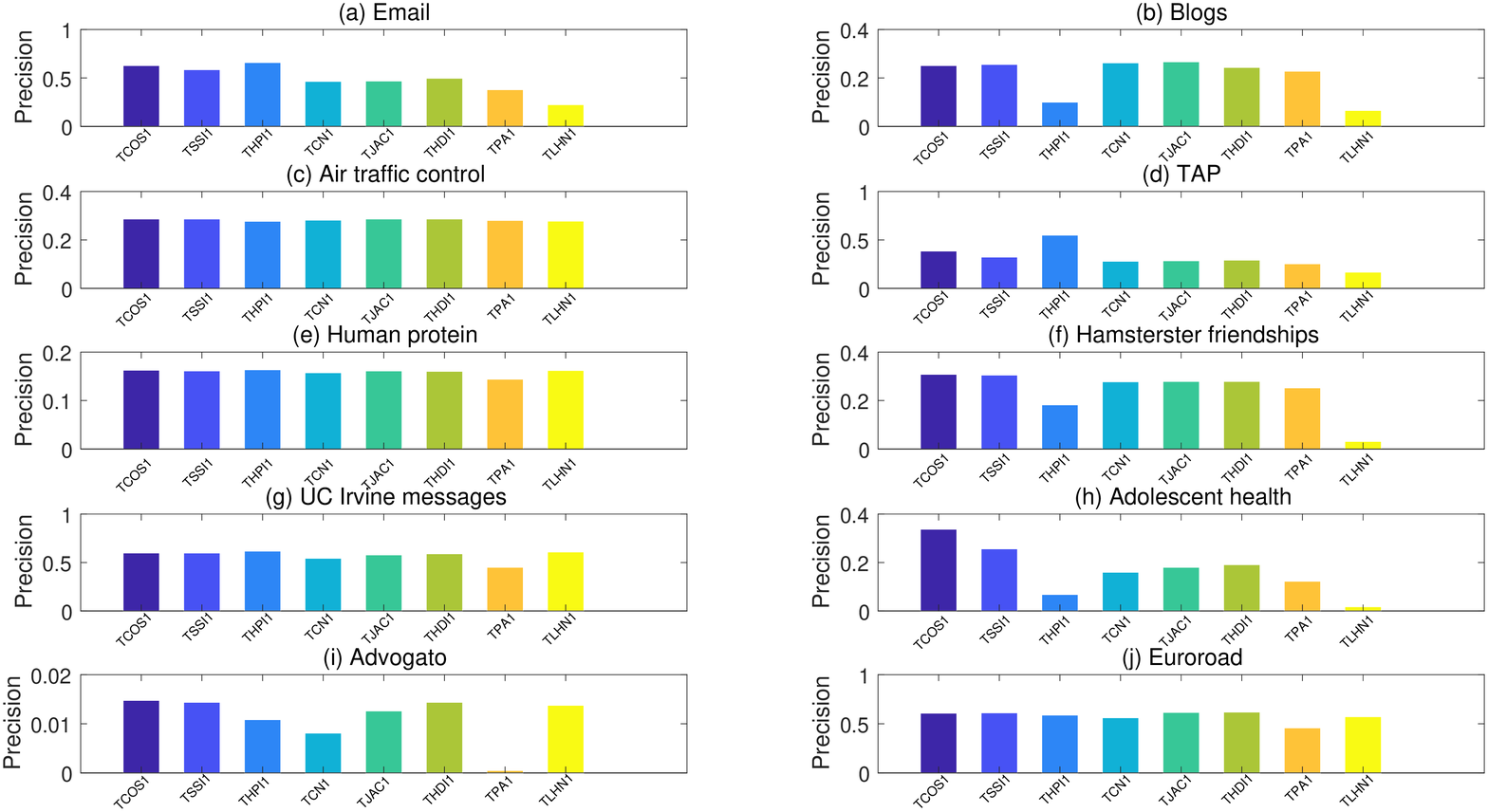}
\caption{Results for the SIR spreading dynamics: Reconstruction precision for $10$ networks. Each panel is the performance of the $8$ TX1 metrics for a single network. This figure shows the results for $\beta = 4 \,\beta_c$, where $\beta_c = \langle k\rangle/(\langle k^2\rangle - \langle k\rangle)$, and $f = 0.5$. The results are averaged over $50$ independent realizations.}
\label{SM5}
\end{figure*}

\begin{figure*}[h!]
  \centering
  \includegraphics[width=17cm,scale=0.5]{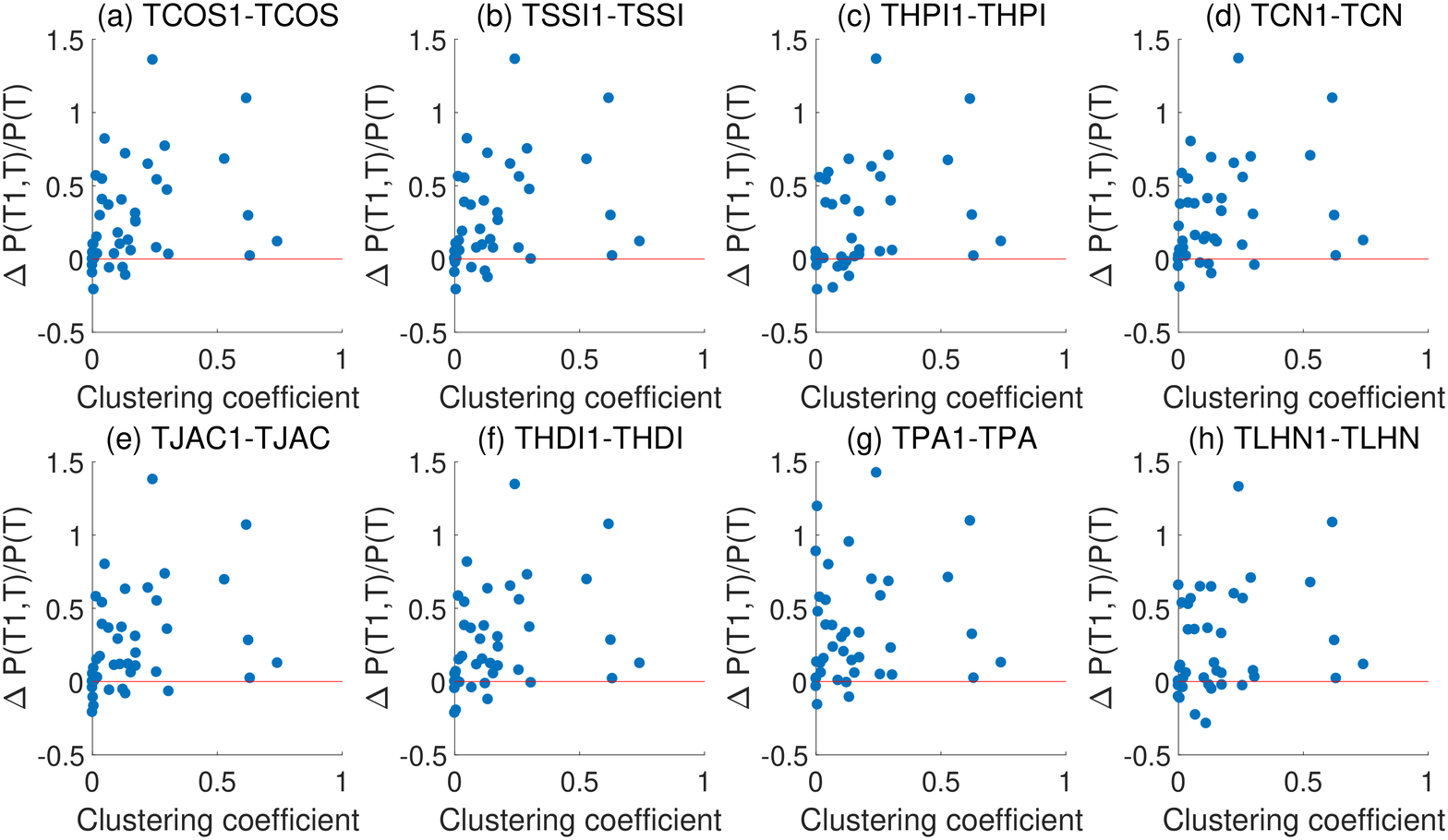}
\caption{Results for the SI spreading dynamics~\citep{anderson1992infectious,starnini2013immunization}.  The SI model is a special case of the SIR model where each infected node cannot recover ($\mu=0$). Reconstruction precision relative difference $\Delta P(T1,T)/P(T)=(P(T1)-P(T))/P(T)$ as a function of the network clustering coefficient. Each dot represents an empirical network; we analyzed $40$ empirical contact networks. For all classes of similarity, almost all the empirical networks fall above the P(T1)=P(T) red line; the only exceptions are some of the networks with low clustering coefficient. We used $\beta = 4 \langle k\rangle/(\langle k^2\rangle - \langle k\rangle)$, and $f = 0.5$ here. The results are averaged over $50$ independent realizations.
}
\label{SM6}
\end{figure*}

\begin{figure*}[h!]
  \centering
  \includegraphics[width=17cm,scale=0.5]{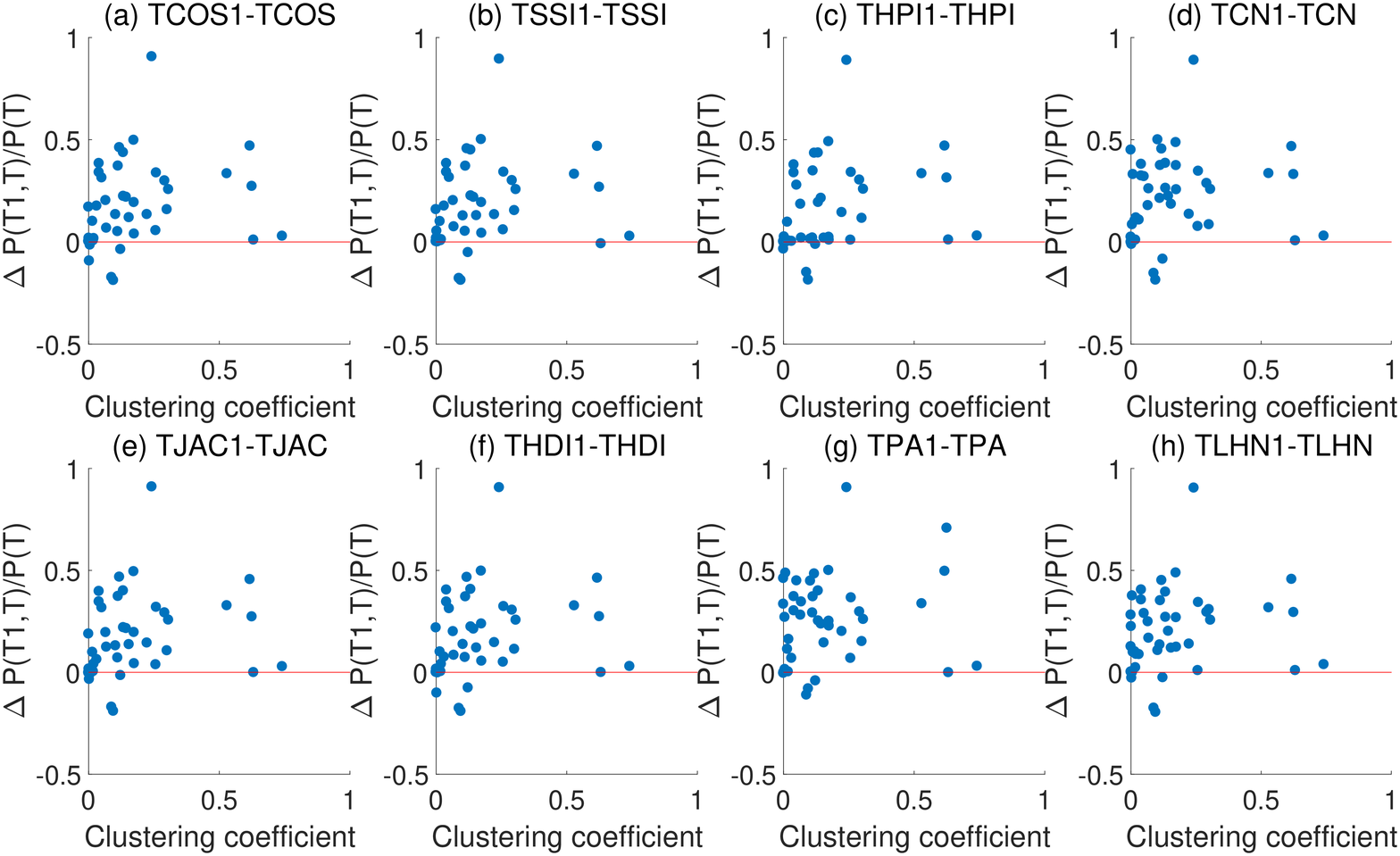}
\caption{Results for the LTM (Linear Threshold Model) \citep{chen2010scalable,granovetter1978threshold}. In the LTM, we replace a link by a pair of directed links, $A \to B$ and $B\to A$, and every directed edge has a weight $1/k_i^{ind}$ , where $k_i^{ind}$ represents the indegree of node $i$. Each node is in one among two possible states, Inactive or Active. A node activates if $\sum_{j}A_{ij}W_{ij}\geq \theta_i$, where $\theta_i$ represents node $i$'s threshold. In our simulations, we start from an initial condition with a fraction $f=0.5$ of randomly-selected Active nodes, and we set $\theta_i=\theta=0.1$. We show here
the reconstruction precision relative difference $\Delta P(T1,T)/P(T)=(P(T1)-P(T))/P(T)$ as a function of the network clustering coefficient. Each dot represents an empirical network; we analyzed $40$ empirical contact networks. For all classes of similarity, almost all the empirical networks fall above the P(T1)=P(T) red line; the only exceptions are some of the networks with low clustering coefficient. The results are averaged over $50$ independent realizations.}
\label{SM7}
\end{figure*}
\newpage

\begin{figure*}[h!]
  \centering
  \includegraphics[width=17cm,scale=0.5]{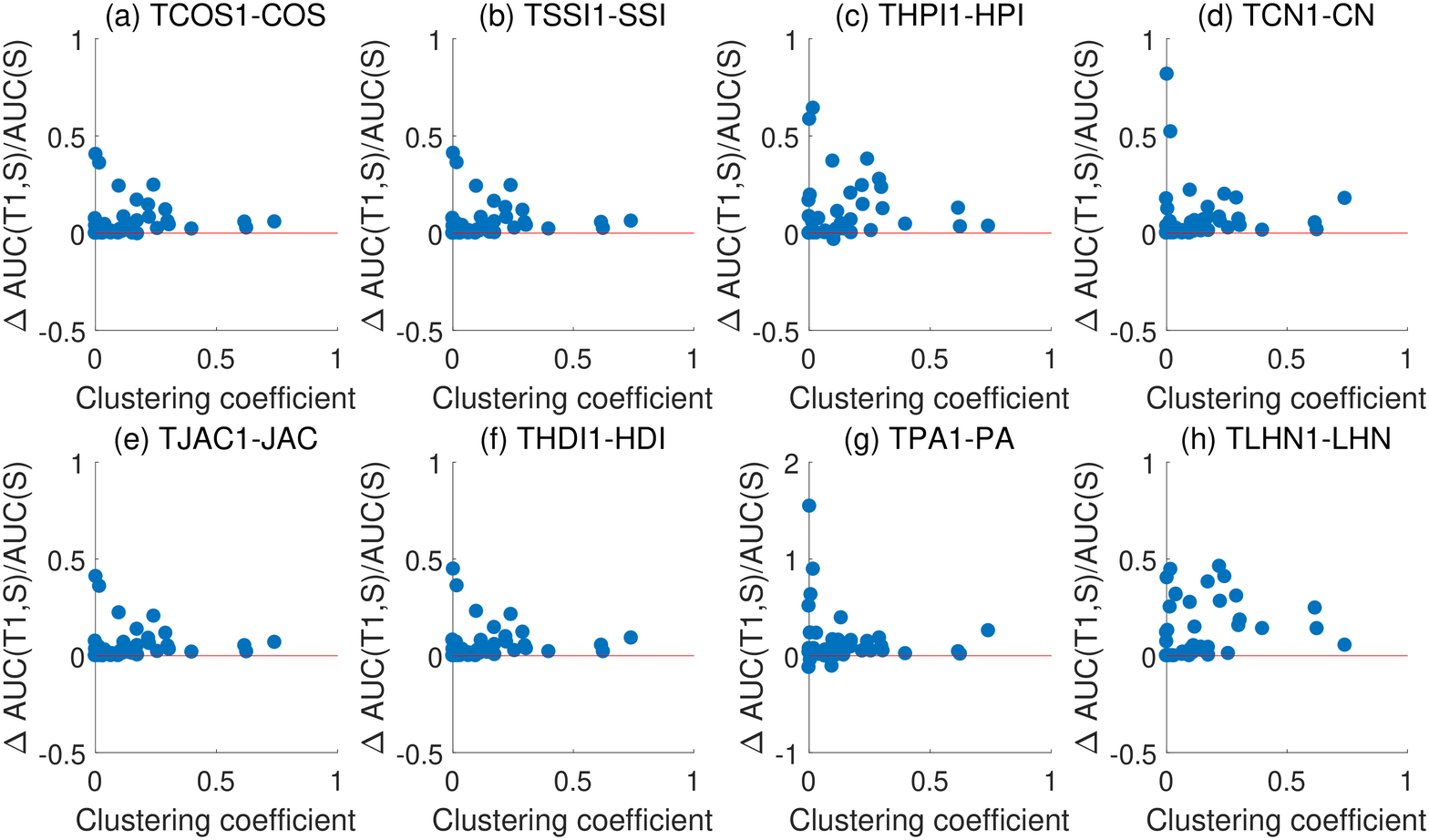}
\caption{Results for the SIR spreading dynamics: Reconstruction AUC relative difference $\Delta AUC(T1,S)/AUC(S)=(AUC(T1)-AUC(S))/AUC(S)$ as a function of the network clustering coefficient (see Methods for the definition). Each dot represents an empirical network, we totally analyzed $40$ empirical network. We use $f = 0.5$, $\beta$ = 4 $\beta_c$ here, where $\beta_c = \langle k\rangle/(\langle k^2\rangle - \langle k\rangle)$. The results are averaged over $50$ independent realizations.}
\label{SM8}
\end{figure*}

\newpage

\newpage
\section*{}



\end{document}